\documentclass[a4paper,11pt]{article}
\usepackage{jheppub}

\usepackage{graphicx,epsfig}
\usepackage{simplewick}
\usepackage{slashed}
\usepackage{color}
\usepackage{setspace}
\usepackage{hyperref}
\usepackage{empheq}

\def\simg{{\ \lower-1.2pt\vbox{\hbox{\rlap{$>$}\lower6pt\vbox{\hbox{$\sim$}}}}\ }}
\def\siml{{\ \lower-1.2pt\vbox{\hbox{\rlap{$<$}\lower6pt\vbox{\hbox{$\sim$}}}}\ }} 

\newcommand{\pert}{\mathrm{pert}}
\newcommand{\order}{\mathcal{O}}

\newcommand{\eq}[1]{Eq.~\eqref{#1}}
\newcommand{\eqs}[2]{Eqs.~\eqref{#1} and \eqref{#2}}

\newcommand{\Sec}[1]{Sec.~\ref{#1}}

   \graphicspath{{./figs/}}
\usepackage{color}

\usepackage{relsize}
\newcommand{\babar}{{\mbox{\slshape B\kern-0.1em{\smaller A}\kern-0.1em
            B\kern-0.1em{\smaller A\kern-0.2em R}}}
\def\MSbar{\relax\ifmmode\overline                        %%%%%%%%%
            {\rm MS}\else{$\overline{\rm MS}${ }}\fi}     %%%%%%%%%
           }                                              %%%%%%%%%
\def\MSbar{\relax\ifmmode\overline                        %%%%%%%%%
            {\rm MS}\else{$\overline{\rm MS}${ }}\fi}     %%%%%%%%%
\def\1{\hbox{{1}\kern-.25em\hbox{l}}}
 \date{\today}
\def\be{\begin{equation}}
\def\ee{\end{equation}}
\def\bea{\begin{eqnarray}}
\def\eea{\end{eqnarray}}
\def\bear{\begin{array}}
\def\eear{\end{array}}

\def\al{\alpha}
\def\nn{\nonumber}

\newcommand{\MS}{\overline{\rm MS}}

\newcommand{\m}{{\overline m}}

\def\lQ{\Lambda_{\rm QCD}}

\begin{document}
\title{Determination of $\alpha(M_z)$ from an hyperasymptotic approximation to the energy of a static quark-antiquark pair}
\author[a]{Cesar Ayala} 
\author[b]{Xabier Lobregat} 
 \author[b]{Antonio Pineda}
\affiliation[a]{Department of Physics, Universidad T{\'e}cnica Federico
Santa Mar{\'\i}a (UTFSM),  
Casilla 110-V, 
Valpara{\'\i}so, Chile}
\affiliation[b]{Grup de F\'{\i}sica Te\`orica, Dept. F\'\i sica and IFAE-BIST, Universitat Aut\`onoma de Barcelona,
E-08193 Bellaterra, Barcelona, Spain}

\date{\today}

\abstract{
We give the hyperasymptotic expansion of the energy of a static quark-antiquark pair with a precision that includes the effects of the subleading renormalon. The terminants associated to the first and second renormalon are incorporated in the analysis when necessary. In particular, we determine the normalization of the leading renormalon of the force and, consequently, of the subleading renormalon of the static potential. We obtain $Z_3^F(n_f=3)=2Z_3^V(n_f=3)=0.37(17)$. The precision we reach in strict perturbation theory is next-to-next-to-next-to-leading logarithmic resummed order both for the static potential and for the force. We find that the resummation of large logarithms and the inclusion of the leading terminants associated to the renormalons are compulsory to get accurate determinations of $\Lambda_{\MS}$ when fitting to short-distance lattice data of the static energy. We obtain $\Lambda_{\MS}^{(n_f=3)}=338(12)$ MeV and $\al(M_z)=0.1181(9)$. We have also found strong consistency checks that the ultrasoft correction to the static energy can be computed at weak coupling in the energy range we have studied. 
}

\maketitle
%\tableofcontents

\vfill
\newpage

\section{Introduction}

The static potential or, more precisely, the energy of a static quark-antiquark pair separated by a distance $r$, is one of the objects most accurately studied by lattice simulations.
This is due to its relevance in order to understand the dynamics of QCD. On the one hand, it
is a necessary ingredient in a Schroedinger-like description of the Heavy Quarkonium dynamics. On the
other hand, a linear growing behavior at long distance is signaled as a proof of confinement.
Moreover, throughout the last years, lattice simulations with dynamical fermions have improved their predictions
at short distances, see for instance \cite{Cheng:2007jq,Bazavov:2012ka,Kaneko:2013jla,Bazavov:2014soa,Bazavov:2017dsy,Bazavov:2018wmo,Karbstein:2018mzo,Weber:2018bam,Bazavov:2019qoo}. In addition, the accuracy of the perturbative prediction of the static
potential has also improved significantly over the years \cite{Fischler:1977yf,Schroder:1998vy,Brambilla:1999qa,Gorishnii:1991hw,Smirnov:2008pn,Anzai:2009tm,Smirnov:2009fh,Pineda:2000gza,Brambilla:2009bi,Pineda:2011db,Brambilla:2006wp,Pineda:2011aw,Lee:2016cgz}. The precision reached nowadays is next-to-next-to-next-to-leading order (NNNLO) for fixed order computation and next-to-next-to-next-to-leading logarithmic (NNNLL) order for renormalization group (RG) improved computations.\footnote{Fixed order computations trivially incorporate the resummation of soft logarithms by setting the renormalization scale to be proportional to $1/r$ up to a coefficient of order one. In this paper the N$^k$LL RG improved expressions refer to those on which the resummation of ultrasoft logarithms is also implemented.} 

The combination of these two items: high order perturbation theory and lattice data at short distances, potentially allows quantitative comparison between perturbation theory and lattice simulations. Nevertheless, naive comparison between lattice data and perturbative results may lead to strong disagreement depending on how the perturbative expansion is implemented in practice. Such different behavior can be understood \cite{Pineda:2002se} on the basis of a renormalon based picture. Overall, analyses that correctly implement the renormalon cancellation (even if it is not mentioned explicitly) show reasonable agreement between lattice data and the corresponding perturbative expressions. This was something that was observed around twenty years ago, see for instance \cite{Necco:2001gh,Recksiegel:2001xq,Pineda:2002se,Lee:2002sn,Brambilla:2009bi,Brambilla:2010pp}. Nowadays, the more recent unquenched data and the knowledge of higher orders in perturbation theory have allowed to obtain competitive determinations of $\Lambda_{\MS}^{(n_f=3)}$, like those in \cite{Bazavov:2014soa,Karbstein:2018mzo,Takaura:2018lpw,Takaura:2018vcy,Bazavov:2019qoo}. Nevertheless, none of them have implemented the resummation of large ultrasoft logarithms to NNNLL order. This is something that we will do in this paper. In particular, expressions for the force with NNNLL precision are given for the first time. 

A most relevant, and novel aspect, of this paper will be the use of hyperasymptotic expansions to deal with the asymptotic behavior associated to the renormalons. Hyperasymptotic expansions were first introduced in the context of the asymptotic solutions to ordinary differential equations \cite{BerryandHowls,Boyd99}. We use here its generalization to quantum field theories with a running coupling constant and renormalons developed in \cite{HyperI,HyperMass,Ayala:2019lak}. Such an approach allows us to compute observables with a well-defined power counting and reach exponential accuracy. In \cite{HyperI}, the general formalism was explained in great detail, and the static potential in the large $\beta_0$ approximation was used as a test-case and analyzed in great detail as well. In \cite{HyperMass} ultraviolet renormalons were explicitly included in the formalism, and the pole mass, and other related objects, were studied in great detail. The general counting for the truncation of the hyperasymptotic approximation was also given in this reference (see also \cite{Ayala:2019lak} where a summary of such counting is given). In this paper, we give the hyperasymptotic expansion of the energy of a static quark-antiquark pair with a precision that includes the effects of the subleading renormalon. The terminants associated to the first and second renormalon are incorporated in the analysis when necessary. In particular, we determine the normalization of the leading renormalon of the force and, consequently, of the subleading renormalon of the static potential. Note that the incorporation of the effects associated to the subleading renormalon of the static potential (which is the leading one of the force) is novel compared with the analyses of \cite{Bazavov:2014soa,Karbstein:2018mzo,Bazavov:2019qoo}. On the other hand \cite{Takaura:2018lpw,Takaura:2018vcy} use a different method to handle the renormalon singularities, and the subleading renormalon has also been studied in the situation when $\alpha/r \ll \Lambda_{\rm QCD}$.

\section{Hyperasymptotic expansion of the static energy}
The energy of a static quark and a static antiquark in a colour singlet configuration separated by a distance $r$, $E(r)$, admits an operator product expansion using pNRQCD \cite{Pineda:1997bj,Brambilla:1999xf}: 
\be
\label{Es1}
E(r)= V(r;\nu_{us})+\delta E_{us}(r;\nu_{us})
\,.
\ee
$ V(r;\nu_{us})$ only encodes the physics associated to the scale $1/r$. $\delta E_{us}(r;\nu_{us})$ encodes the physics associated to scales smaller than $1/r$ and appears at ${\cal O}(r^2)$ in the multipole expansion. 

As explained at length in \cite{HyperI} we need to give meaning to the different terms of the operator product expansion with nonperturbative 
power accuracy. In particular this implies that we have to regularize the perturbative series expansion of $V$. We define it using the principal value (PV) prescription, i.e. the average of the Borel integrals infinitesimally above and below the cut in the real plane. See \cite{Dingle} and the appendix of \cite{HyperI} for explicit expressions. We then have to use the same prescription for $\delta E_{us}$. Therefore, we have 

\be
E^{\rm PV}(r)=V_{\rm PV}(r;\nu_{us})+\delta E_{us}^{\rm PV}(r;\nu_{us})
\,.
\ee
The asymptotic expansion of $V_{\rm PV}$ coincides, by construction, with the asymptotic expansion of $V$ and reads
\be
V_{\rm PV}(r;\nu_{us}) \sim \sum_{n=0}^{\infty} V_n \al^{n+1}(\nu_s)
\,.
\ee
The natural scale for this expansion is $\nu_s \sim 1/r$. The coefficients 
\be
\label{eq:Vn}
V_n \equiv 
 -\frac{C_F}{r}\left(\frac{1}{4\pi}\right)^n a_n(\nu_s r;\frac{\nu_{us}}{\nu_s}) 
\ee 
were computed in \cite{Fischler:1977yf,Schroder:1998vy,Brambilla:1999qa,Smirnov:2008pn,Anzai:2009tm,Smirnov:2009fh} for $n=0,1,2,3$.  For ease of reference we quote them in Appendix \ref{constants}.

The perturbative expansions of the pole mass, and of the static potential in the large $\beta_0$ approximation, are characterized by having a single scale (leaving aside $\lQ$): the heavy quark mass renormalized in the $\MS$ scheme, $\m$, for the case of the pole mass, and $1/r$ for the case of the static potential in the large $\beta_0$ approximation. In other words, they are infrared finite at any finite order in perturbation theory. This is not so for $V_{\rm PV}$. This reflects in that $V_{\rm PV}$ is logarithmically  infrared divergent. Such behavior first appears in $V_3$ (see \eq{eq:Vr}), which endures a linear $\ln (\nu_{us})$ dependence that is not absorbed in the renormalization of the strong coupling in $V_{\rm PV}$. This logarithmic behavior is cancelled instead by $\delta E_{us}$. Therefore, both $V_{\rm PV}(\nu_{us})$ and $\delta E_{us}(\nu_{us})$, unlike the pole mass, are renormalization scale and scheme dependent in perturbation theory. Such scale and scheme dependence cancels in the sum in \eq{Es1}. 

Powers of $\al(\nu_s)\ln (\frac{\nu_s}{\nu_{us}})$ (we take $\nu_s \sim 1/r$) can be resummed using RG techniques. They have been computed in \cite{Pineda:2000gza} with leading logarithmic (LL) accuracy and in \cite{Brambilla:2009bi} with next-to-leading logarithmic (NLL) accuracy (see also \cite{Pineda:2011db}). This produces the following correction to $V_{\rm PV}$:
\be
\label{eq:deltaVRG}
\delta V_{RG}(r;\nu_s,\nu_{us})
=
{\bf r}^2(\Delta V)^3
G(\nu_s;\nu_{us})
\,,
\ee
where ($V_o$ is the potential of a static quark-antiquark pair in a colour octet configuration, and $\beta_0=11/3C_A-4/3T_Fn_f$) 
\be
\Delta V=V_o-V \equiv {C_A
\over 2}{\alpha(\nu_s) \over r}\left(1+\frac{\al}{4\pi}(a_1+2\beta_0\ln(\nu_s e^{\gamma_E} r))+{\cal O}(\al^2)\right)
\,.
\ee
and
\bea
G(\nu_s;\nu_{us})
&=&C_FV_A^2
\frac{2\pi}{\beta_0}
\left\{
\frac{2}{3\pi}\ln\frac{\al(\nu_{us})}{\al(\nu_s)}
\right.
\\
\nn
&&
\left.
-(\al(\nu_{us})-\al(\nu_s))
\left(
\frac{8}{3}\frac{\beta_1}{\beta_0}\frac{1}{(4\pi)^2}-\frac{1}{27\pi^2}\left(C_A\left(47+6\pi^2\right)-10T_Fn_f\right)
\right)
\right\}
\\
\nn
&
\simeq& -C_F V_A^2\frac{2}{3}\frac{\al}{\pi}\ln \frac{\nu_{us}}{\nu_s}
\,,
\eea
where $V_A=1$ with LL \cite{Pineda:2000gza}
and NLL \cite{Brambilla:2009bi} accuracy. 
Then, the RG improved potential reads
\be
V_{\rm PV}^{\rm RG}=V_{\rm PV}(r;\nu_{us}=\nu_s)+\delta V_{\rm RG}(r;\nu_s,\nu_{us})
\,.
\ee
$\delta E_{us}(r;\nu_{us})$ encodes the physics associated to scales smaller than $1/r$ and can be computed using the multipole expansion. At ${\cal O}(r^2)$ its explicit expression reads (in the Euclidean)
\be
\label{energyUS}
\delta E_{us}(r;\nu_{us})= {T_F \over 3 N_c}  {\bf r}^2 V_A^2
\int_0^\infty \!\! dt 
 e^{-t \Delta V} \langle g{\bf E}^a(t) 
\phi(t,0)^{\rm adj}_{ab} g{\bf E}^b(0) \rangle(\nu_{us}).
\ee

This quantity has a different behavior depending on the relative size between $\lQ$ and $\Delta V \sim \al(\nu_s)/r$. If $\lQ \gg  \al(\nu_s)/r$ the above expression can be approximated to 
 \be
\label{dEusr2}
\delta E_{us}(r;\nu_{us})= {T_F \over 3 N_c} {\bf r}^2 V_A^2
\int_0^\infty \!\! dt 
 \langle g{\bf E}^a(t) 
\phi(t,0)^{\rm adj}_{ab} g{\bf E}^b(0) \rangle \sim r^2\lQ^3. 
\ee
On the other hand, if $\lQ \ll \Delta V$, $\delta E_{us}$ can be computed at weak coupling as an expansion in powers of $\al(\nu_{us})$. It has the following scaling 
\be
\delta E_{us}(r;\nu_{us}) \sim r^2(\Delta V)^3H(\al(\nu_{us}))
\,,
\ee
where $\Delta V$ is generated dynamically and $H(\al(\nu_{us}))$ admits a perturbative expansion in powers of $\al(\nu_{us})$ (up to logarithms). $\delta E_{us}(\nu_{us})$ is known to order $r^2(\Delta V)^3 \al^2(\nu_{us}) \sim \frac{1}{r}\al^5$  in the $\MS$ scheme:\footnote{Its expression in the large $\beta_0$ approximation can be found in \cite{Sumino:2004ht}.} 
\be
\label{eq:deltaEUS}
\delta E_{us}^{\MS}|_{\rm LO}
=
-C_F {\bf r}^2(\Delta V)^3 V_A^2
\frac{\al(\nu_{us}) }{9 \pi } \left(6\ln\frac{\Delta V}{\nu_{us}}+6\ln 2-5\right)
\,,
\ee
\bea
&&
\delta E_{us}^{\MS}|_{\rm NLO}
=
C_F {\bf r}^2(\Delta V)^3 V_A^2
\frac{\al^2(\nu_{us}) }{108 \pi ^2} 
\left(18 \beta_0 \ln ^2\left(\frac{\Delta V}{\nu_{us}
   }\right)
\right.
\\
\nn
&&
-6 \left(C_A \left(13+4 \pi ^2\right)-2 \beta_0 (-5+3\ln 2)\right) \ln \left(\frac{\Delta V}{\nu_{us}
   }\right)
   \\
   &&
   \nn
-2 C_A \left(-84+39 \ln 2+4 \pi ^2 (-2+3\ln 2)+72 \zeta (3)\right)+\beta_0 \left(67+3 \pi ^2-60 \ln 2+18\ln^2 2\right)\Bigg)
\,,
\eea 
where we take the next-to-leading order (NLO) expression from \cite{Pineda:2011aw}.
This is one order more than we have for $V_{\rm PV}$: $\sim \frac{1}{r}\al^4$. We remind again that $\delta E_{us}$ is scheme dependent.

Finally, if both scales, $\lQ$ and $\Delta V$, are similar in size, $\delta E_{us}(r;\nu_{us})$ is an unknown function of the ratio of these two scales. 

We now focus on the hyperasymptotic approximation to $V_{\rm PV}$. 
The leading asymptotic behavior of $V_n$ is known (and up to a factor minus two is equal to the asymptotic behavior of the pole mass \cite{Pineda:1998id}). It reads
\be
\label{rnas}
V_n^{\rm (as)}(\mu)=Z_{1}^{V} \mu \,\left({\beta_0 \over
2\pi}\right
)^n \,\sum_{k=0}^\infty c_k{\Gamma(n+1+b-k) \over
\Gamma(1+b-k)}
\,,
\ee
where $b=\beta_1/(2\beta_0^2)$. 
The coefficients $c_k$ are pure functions of the $\beta$-function coefficients, as first shown in \cite{Beneke:1994rs} for the case of the pole mass. They can be found in \cite{Beneke:1998ui,Pineda:2001zq,Ayala:2014yxa}. At low orders they read ($c_0=1$) 
\be
c_1=s_1 \,,\quad 
c_2=\frac{1}{2}\frac{b}{b-1}(s_1^2-2s_2)
\,,
\quad
c_3=\frac{1}{6}\frac{b^2}{(b-2)(b-1)}(s_1^3-6s_1s_2+6s_3)
\,,
\ee
 where the $s_n$ are defined in \cite{HyperI}. 
 
We construct the hyperasymptotic expansion of $V_{\rm PV}$ along the lines of \cite{HyperI,HyperMass}. It does not have ultraviolet renormalons, whereas the leading infrared ones are located at dimension 1 and 3 (i.e. at $u=1/2$ and at $u=3/2$ in the Borel plane). The termination of the perturbative series associated to these renormalons produces nonperturbative power contributions of order $\lQ$ and $r^2 \lQ^3$ respectively:
\be
\label{eq:VPV}
V_{\rm PV}=V_{P}+\frac{1}{r}\Omega_1^V+\sum_{n=N_P+1}^{3N_P} (V_n-V_n^{(\rm as)}) \al^{n+1}(\nu_s)
+\frac{1}{r}\Omega_3^V+o(\lQ^3 r^2)
\,,
\ee
where
\be
V_P\equiv \sum_{n=0}^{N_P} V_n \al^{n+1}(\nu_s)\; ;
\ee 
and
\be
\label{eq:NP}
N_P=\frac{2\pi}{\beta_0\al(\nu_s)}\left(1-c\al(\nu_s)\right)
\,.
\ee
Approximating $V_{\rm PV}$ by $V_{\rm P}$ corresponds to achieve superasymptotic approximation. In the generic labeling $(D,N)$ of the truncations of the hyperasymptotic approximation defined in \cite{HyperMass,Ayala:2019lak}, it corresponds to $(0,N_P)$ precision. The next order in the hyperasymptotic approximation is labeled as (1,0) and means adding $\Omega_1^V/r$ to $V_P$. Its explicit expression reads (in \cite{HyperI} $\Omega_1^V$ was named $\Omega_V$, and $Z_1^V$ was named $Z_V$)
\begin{equation}
\label{eq:OmegaV}
\Omega_1^V=\sqrt{\alpha(\nu_s)}K^{(P)}
\nu_s \, r
e^{-\frac{2\pi}{\beta_0 \alpha(\nu_s)}}
\left(\frac{\beta_0\alpha(\nu_s)}{4\pi}\right)^{-b}
\bigg(1+\bar K_{1}^{(P)}\alpha(\nu_s)+\bar K_{2}^{(P)}\alpha^2(\nu_s)+\mathcal{O}\left(\alpha^3(\nu_s)\right)\bigg)
\,,
\end{equation}
\bea
K^{(P)}&=&-\frac{Z_1^V 2^{1-b}\pi}{\Gamma(1+b)}\beta_0^{-1/2}\bigg[-\eta_c+\frac{1}{3}\bigg]
\,,
\\
\bar K_{1}^{(P)}&=&\frac{\beta_0/(\pi)}{-\eta_c+\frac{1}{3}}\bigg[-b_1 b \left(\frac{1}{2}\eta_c+\frac{1}{3}\right)
-\frac{1}{12}\eta_c^3+\frac{1}{24}\eta_c-\frac{1}{1080}\bigg]
\,,
\\
\bar K_{2}^{(P)}&=&\frac{\beta_0^2/\pi^2}{-\eta_c+\frac{1}{3}}
\bigg[-w_2 (b -1) b \left(\frac{1}{4}\eta_c+\frac{5}{12}\right)
+b_1b\left(-\frac{1}{24}\eta_c^3-\frac{1}{8}\eta_c^2
-\frac{5}{48}\eta_c-\frac{23}{1080}\right)
\nn
\\
&&
-\frac{1}{160}\eta_c^5
-\frac{1}{96}\eta_c^4+\frac{1}{144}\eta_c^3
+\frac{1}{96}\eta_c^2-\frac{1}{640}\eta_c-\frac{25}{24192}\bigg]
\,,
\eea
and so on, where we have applied the general expression obtained in \cite{HyperI} to this case.  In particular,
\be
\eta_c=-b d+\frac{2\pi d  c}{\beta_0}-1 \;, \quad b_1=d s_1 \;, \quad {\rm and} \quad w_2=\left(\frac{s_1^2}{2}-s_2\right)\frac{b}{b-1}
\,,
\ee
where for $w_2$ we have already set $d=1$ for simplicity. 
 
Note that the $u=1/2$ renormalon does not cancel with the renormalons in $\delta E_{us}$. \eq{eq:OmegaV} exactly corresponds to $-2\Omega_m$ defined in \cite{HyperMass}, since the renormalon of twice the pole mass cancels with the renormalon of the static potential (in other words $Z_1^V=-2Z_m$). Only after the inclusion of the pole mass in real observables, this renormalon cancels. This is the reason we have to specify the prescription used to regularize the perturbative sum of $E$. Had we included $2 m_{\rm OS}$, the summation scheme dependence would have disappeared. 

We now move beyond (1,0) hyperasymptotic precision by including the third term in \eq{eq:VPV}. One then generically reaches hyperasymptotic precision (1,$N$). For $N$ large one would start to be sensitive to the next renormalon, which then has to be considered. 
We can use \eq{dEusr2} to determine the renormalon structure of the subleading infrared renormalon. 
Due to the fact that $V_A=1$ with NLL accuracy, we can determine the first two terms of the asymptotic expansion of the perturbative series associated to the $u=3/2$ renormalon of the static potential (we can do similarly for the force, see \eq{fpn}).
 The best way to quantify the asymptotic behaviour of the perturbative series is by
performing its Borel transform:
\be
B[rV_{\pert}]\equiv\sum_{n=0}^{\infty}\frac{rV_n}{n!}\left(\frac{4\pi}{\beta_0}u\right)^{\!n}\,.
\ee
The Borel transform will have a singularity, due to the dimension $d=3$ non-local condensate, at $u=d/2=3/2$:
\be
\label{eq:borel}
B[rV_{\pert}]
\dot=
Z_3^{V}(r \mu)^3 \, \frac{1}{(1-2u/d)^{1+db}}\left[1+b_1\left(1-\frac{2u}{d}\right)+\cdots\right]\,,
\ee
where $b_1=ds_1$. This singularity produces the following asymptotic behavior\footnote{This and the previous expression have also been presented in \cite{Sumino:2020mxk} at leading order. Nevertheless, they do not include the subleading corrections we have presented here, as they argue, differently to what conclude in this paper, that they are not known.}
\begin{align}
\label{pn}
r(V_n-V_n^{(\rm as)}) &\stackrel{n\rightarrow\infty}{=} Z_3^V(r \mu)^3\,
\left(
\frac{\beta_0}{2\pi d}\right)^{\!n}
\frac{\Gamma(n+1+db)}{\Gamma(1+db)}
\left\{
1+\frac{db}{n+db}\,b_1
+
\order\left(\frac{1}{n^2}\right)
\right\}
\,.
\end{align}
This asymptotic behavior has associated the terminant $\Omega_3^V$, which reads
\begin{equation}
\label{eq:OmegaVprime}
\Omega_3^V=\sqrt{\alpha(\nu_s)}K^{'(P)}
r^3\nu^3_s
e^{-3\frac{2\pi}{\beta_0 \alpha(\nu_s)}}
\left(\frac{\beta_0\alpha_X(\nu_s)}{4\pi}\right)^{-3b}
\bigg(1+\bar K_{1}^{'(P)}\alpha(\nu_s)+\mathcal{O}\left(\alpha^2(\nu_s)\right)\bigg)
\,,
\end{equation}
where
\bea
K^{'(P)}&=&-\frac{Z_3^V 2^{1-3b}\pi 3^{3 b+1/2}}{\Gamma(1+3b)}\beta_0^{-1/2}\bigg[-\eta_c+\frac{1}{3}\bigg]
\,,
\\
\bar K_{1}^{'(P)}&=&\frac{\beta_0/(3\pi)}{-\eta_c+\frac{1}{3}}\bigg[-3 b_1 b \left(\frac{1}{2}\eta_c+\frac{1}{3}\right)
-\frac{1}{12}\eta_c^3+\frac{1}{24}\eta_c-\frac{1}{1080}\bigg]
\,.
\eea

If we have enough terms in perturbation theory to be sensitive to the $u=3/2$ renormalon, we may reach order $(3,0)$ precision in the hyperasymptotic counting for $V_{\rm PV}$ (see \cite{HyperMass,Ayala:2019lak} for a detailed account of the hyperasymptotic counting we use in this paper). This is the maximal accuracy we will test in this paper. 

In principle, we will work under the hypothesis that we have enough terms in perturbation theory to be sensitive to the $u=3/2$ renormalon and that $\lQ \ll \al/r$ so that $\delta E^{\rm PV}_{us}$ can be computed in perturbation theory, though we will also test predictions where $\delta E^{\rm PV}_{us}$ is modeled by a $\lQ^3 r^2$ term.
In any case, the final expression we obtain reads
\be
\label{eq:Esapprox}
E^{\rm PV}=V_{\rm PV}^{\rm RG}(r;\nu_s,\nu_{us})+ \delta E^{\rm PV}_{us}(r;\nu_{us})
\,,
\ee
where 
\bea
\nn
V_{\rm PV}^{\rm RG}(r;\nu_s,\nu_{us})&=&V_{P}(\nu_{us}=\nu_s)+\frac{1}{r}\Omega_1^V+\sum_{n=N_P+1}^{N_{max}} (V_n(\nu_{us}=\nu_s)-V_n^{(\rm as)}) \al^{n+1}\\
&&
+\frac{1}{r}\Omega_3^V
+\delta V_{\rm RG}(r;\nu_s,\nu_{us})
\,,
\eea
and
$N_{max}=3$. 

We will confront the above theoretical expression with nonperturbative evaluations of $E(r)$. 
$E(r)$ can be determined accurately using Montecarlo simulations in the lattice of the quantity
\be
\label{Es}
E^{latt}(r)=\lim_{T \rightarrow \infty}\frac{i}{T}\ln \langle W_\Box \rangle
\,,
\ee
where $W_\Box$ is the rectangular Wilson loop with edges $x_1 = (T/2,{\bf r}/2)$, $x_2 = (T/2,-{\bf r}/2)$, 
$y_1 = (-T/2,{\bf r}/2)$ and  $y_2 = (-T/2,-{\bf r}/2)$. 
The symbol $\langle ~~ \rangle$ means the average over the massless gluons and the light quarks. This quantity is linearly divergent in $1/a$ by an $r$ independent constant. Therefore $E^{\rm PV}(r)$ and $E^{latt}(r)$ are equal up to an additive $r$-independent constant (and up to ${\cal O}(a)$ lattice artefacts which can be $r$ dependent). Leaving aside these lattice artifacts, $E^{\rm PV}(r)$ and $E^{latt}(r)$ have the same $r$ dependence. 

\section{Hyperasymptotic expansion of the force}
\label{Sec:Force}
We now consider the force, which we define as the derivative of the static potential:
\be
F_{\rm PV}(r;\nu_{us})\equiv \frac{d}{dr}V_{\rm PV}(r;\nu_{us}) \sim \sum_{n=0}^{\infty} f_n(\nu_s r) \al^{n+1}(\nu_s)
\,,
\ee
where the present known values of $f_n$ read (we take $\nu_s=x_s/r$, and for $n \geq 3$, $f_n$ also depends on $\nu_{us}$: $f_n(\nu_s r;\frac{\nu_s}{\nu_{us}}$))
\bea
\label{eq:fn}
f_0(x_s)&=&\frac{C_F}{r^2} \;, \qquad f_1(x_s)=\frac{C_F}{4\pi r^2}\left(a_1(x_s)-2\beta_0\right)\,,
\\
\nn
f_2(x_s)&=&\frac{C_F}{(4\pi)^2 r^2}
\left(
a_2(x_s)-4a_1(x_s)\beta_0-2\beta_1
\right)
\,,
\\
\nn
f_3(x_s;\frac{x_s}{r\nu_{us}})&=&\frac{C_F}{(4\pi)^3 r^2}
\left(
a_3(x_s;\frac{x_s}{r\nu_{us}})-6a_2(x_s)\beta_0-4a_1(x_s)\beta_1-2\beta_2-\frac{16}{3}C_A^3\pi^2
\right)
\,.
\eea

$F_{\rm PV}$ admits a strict perturbative expansion in powers of $\alpha(\nu_s)$ (the dependence in $\nu_{us}$ is hidden in the coefficients $f_n$). On the other hand $F_{\rm PV}(r;\nu_{us})$ is not RG invariant, since it is dependent on $\nu_{us}$.  

We also consider the quantity 
\be
\label{eq:calF}
{\cal F}(r)
\equiv
\frac{d}{dr}E_{\rm PV}(r)=F_{\rm PV}(r;\nu_{us})+ \frac{d}{dr}\delta E_{us}^{\rm PV}(r;\nu_{us})
\,,
\ee
where (at leading order and assuming $\lQ \ll \Delta V$)
\be
\frac{d}{dr}\delta E_{us}^{\rm PV}(r;\nu_{us})
=
C_F r(\Delta V)^3 
\frac{\al(\nu_{us}) }{9 \pi} \left(6\ln\frac{\Delta V}{\nu_{us}}+6\ln 2+1\right)
\,.
\ee
If we neglect renormalons, ${\cal F}(r)$ is now known with N$^3$LO precision. Adding all the terms the resulting expression is equal to Eq. (10) of \cite{Bazavov:2014soa}. Notice, though, that the definition of the coefficient $a_3$ is different here and in that paper. This is compensated by the last term in Eq. (\ref{eq:calF}) which is also different. 

We can also make a RG improved version for the force and for ${\cal F}(r)$: 
\be
F^{\rm RG}_{\rm PV}(r;\nu_{us})=F_{\rm PV}(r;\nu_{us}=\nu_s)+\frac{d}{dr}\delta V_{\rm RG}(r;\nu_s,\nu_{us})\,,
\ee
\be
{\cal F}^{\rm RG}(r)=\frac{d}{dr}E^{\rm RG}_{\rm PV}(r)=F_{\rm PV}^{\rm RG}(r;\nu_{us})+ \frac{d}{dr}\delta E_{us}^{\rm PV}(r;\nu_{us})
\,,
\ee
where
\bea
\nn
&&
\frac{d}{dr}\delta V_{\rm RG}(r;\nu_s,\nu_{us})
=-r(\Delta V)^3
G(\nu_s;\nu_{us})+C_FV_A^2 r(\Delta V)^3 \frac{2\al(\nu_s)}{\pi}\ln \frac{\al(\nu_{us})}{\al(\nu_s)}+{\cal O}(\al^5)
\\
\label{eq:derivativedeltaVRG}
&&=
-r\left(\frac{C_A\alpha(\nu_s)}{2r}\right)^3C_F
\left\{
\frac{4}{3\beta_0}
\ln\frac{\al(\nu_{us})}{\al(\nu_s)}
\right.
\\
\nn
&&
+\frac{\al(\nu_s)}{\pi}(a_1+2\beta_0\ln(\nu_s e^{\gamma_E} r))
\frac{1}{\beta_0}\ln\frac{\al(\nu_{us})}{\al(\nu_s)}
- \frac{2\al(\nu_s)}{\pi}\ln \frac{\al(\nu_{us})}{\al(\nu_s)}
\\
\nn
&&
\left.
-\frac{2\pi}{\beta_0}
(\al(\nu_{us})-\al(\nu_s))
\left(
\frac{8}{3}\frac{\beta_1}{\beta_0}\frac{1}{(4\pi)^2}-\frac{1}{27\pi^2}\left(C_A\left(47+6\pi^2\right)-10T_Fn_f\right)
\right)
\right\}+{\cal O}(\al^5)
\,.
\eea
The pure perturbative expression for ${\cal F}^{\rm RG}(r)$ is known at the
NNNLL level. Our expression corrects Eq. (11) of \cite{Bazavov:2014soa} at
the NNNLL level. We emphasize though that such Eq. (11) was not used for
analyses in this reference but rather the same expression we have obtained here\footnote{We thank Xavier Garcia i Tormo for checking this.}. 

Finally, let us emphasize that there is no unique way to decompose ${\cal F}^{\rm RG}(r)$. In the above result we have taken $\nu_s$ and $\nu_{us}$ to be $r$-independent in the derivative, though to resum the large ultrasoft logarithms we have to take $\nu_s \sim 1/r$ and $\nu_{us} \sim \frac{C_A\al(\nu_s)}{2r}$. We can make this dependence explicit: 
$\nu_s=x_s/r$ and $\nu_{us}=x_{us}\frac{C_A\al(\nu_s)}{2r}$. This introduces an explicit dependence on $r$ in $\nu_s$ and $\nu_{us}$. We show how the different terms of ${\cal F}^{\rm RG}(r)$ look like in this situation in Appendix \ref{app:nu}.

We now introduce in the discussion renormalon effects. 
$F_{\rm PV}$ does not have the renormalon at $u=1/2$. The leading renormalon is located at $u=3/2$. On the other hand, 
${\cal F}(r)$ can be considered to be an observable. The uncancelled renormalon at $u=1/2$ that exists in $E_{\rm PV}(r)$ vanishes in ${\cal F}$ after taking the derivative with respect to $r$. All other renormalons of $V_{\rm PV}(r;\nu_{us})$ cancel with the analogous renormalons of $\delta E_{us}^{\rm PV}(r;\nu_{us})$ in $E_{\rm PV}(r)$ and consequently the same cancellation takes place in ${\cal F}(r)$. 

The perturbative expansions of $V$ and $\delta E_{us}$ are series in powers of $\al$ evaluated at different scales: $\al(\nu_s)$ and $\al(\nu_{us})$ respectively. The same thing applies to $F$ and $\frac{d}{dr} \delta E_{us}$. This makes that there is no renormalon cancellation order by order in $\al$. If the perturbative series reaches orders high enough to be sensitive to the $u=3/2$ renormalon, we should indeed incorporate the associated nonperturbative contribution to the PV summation, i.e. the corresponding terminants. The complete expression then reads
\bea
{\cal F}^{\rm RG}(r)=F_{P}(\nu_{us}=\nu_s)
+\frac{d}{dr}\delta V_{\rm RG}(r;\nu_s,\nu_{us})+\frac{d}{dr}\delta E_{us}^{P}(r;\nu_{us})+
\frac{1}{r^2}\Omega^F_{3}(\nu_s)-\frac{1}{r^2}\Omega^F_{3}(\nu_{us})
\,,
\nn
\\
\label{eq;calFRG}
\eea
where 
\be
F_{P}(r;\nu_{us}=\nu_s)=\sum_{n=0}^{N_F} f_n \al^{n+1}(\nu_s)
\ee
 and 
 \be
 \frac{d}{dr}\delta E_{us}^{P}(r;\nu_{us})=\sum_{n=0}^{N_{us}} p_n \al^{n+1}(\nu_{us})
 \ee
are the superasymptotic approximations of $F_{\rm PV}$ and $\frac{d}{dr}\delta E_{us}^{\rm PV}(r;\nu_{us})$. For them we have 
$
N_F=3\frac{2\pi}{\beta_0\al(\nu_s)}\left(1-c_F\al(\nu_s)\right)$ and 
$
N_{us}=3\frac{2\pi}{\beta_0\al(\nu_{us})}\left(1-c_{us}\al(\nu_{us})\right)$.
We will usually take $N_F=3$ and fine tune the coefficient $c_F$ accordingly. For $N_{us}$ we know less orders of the perturbative expansion, and the scale $\nu_{us}$ is small. Therefore, 
we will usually take $N_{us}=0$ and fine tune the coefficient $c_{us}$ accordingly.

As $N_F$ is small, there is some degree of fine-tuning between the relative size of 1 and $c_F\alpha$. This is not a problem because to determine whether the expansion in the terminant in \eq{eq:OmegaVprime} makes sense, we do not have to look to $c_F$ but rather to the size of the complete correction. In other words, to see whether the subleading term $K'_1\alpha_X$ is smaller than one (except if, for some reason, the leading order is anomalously small). In the range of $\nu_s$ we cover with the data set I we use in \Sec{Sec:Fit}, this term is in the range
$-0.055 < \bar K_{X,1}^{'(P)}\alpha_X(\nu_s) < 0.11$. Therefore, it is safely small. Even for the ultrasoft case, where the situation is potentially worse, we find $-0.29 < \bar K_{X,1}^{'(P)}\alpha_X(\nu_{us}) < 0.0006$
for the range of scales of the data set I. We find again that the correction is safely smaller than one.
As an extreme test, we could even set this subleading correction to zero (set $ \bar K_{X,1}^{'(P)}=0$). We find that the fit shifts by 2 MeV only. On top of that one could be worried about a possible asymptotic character of the weak coupling expansion of \eq{eq:OmegaVprime}. At present we cannot make definite statements about the possible asymptotic nature of this expansion (see also the discussion in Appendix A of \cite{HyperMass} and \cite{Dingle}) but from the previous numerical discussion we do not see signs of divergence of the weak coupling expansion. 

The last items of \eq{eq;calFRG} are the terminants of the soft and ultrasoft perturbative series. For them we have 
$\frac{1}{r^2}\Omega_3^F=\frac{d}{d r} \frac{1}{r}\Omega_3^V$. Notice that $\Omega_3^F(\nu_s)$ and $\Omega_3^F(\nu_{us})$ are different, not only because of the different renormalization scale each of them uses, but also because we truncate the perturbative expansion at different orders in $F$ and $\frac{d}{dr} \delta E_{us}$. Therefore, even if we set $\nu_{us}=\nu_s$, the terminants will not cancel each other in general (they would only do if we truncate the perturbative expansions of $F$ and $\frac{d}{dr} \delta E_{us}$ to the same order). 

\section{Normalization of the $u=3/2$ renormalon}
\label{u}

The hyperasymptotic expressions derived in the previous sections are completely determined up to the normalizations of the $u=1/2$ and $u=3/2$ renormalons: $Z_1^V$ and $Z_3^V$ respectively. They can only be computed approximately. For $Z_1^V$ we use the value determined in \cite{Ayala:2014yxa} using the static potential: $Z_{1,\MS}^V= -1.1251(520)$. The direct determination of $Z_3^V$ from $V$ is complicated because of the $u=1/2$ renormalon. On the other hand, the force is not contaminated by the $u=1/2$ renormalon. Therefore, it is an ideal place where to see if the perturbative series, as we know it at present, is sensitive to the subleading (infrared) renormalon, which is located at $u=3/2$ in the Borel plane. 

\begin{figure}[htb] %\unitlength=1mm
\begin{minipage}[b]{.49\linewidth}
 %\centering\includegraphics[width=85mm]{RatioForceCoeffNf3.eps}
  \centering\includegraphics[width=80mm]{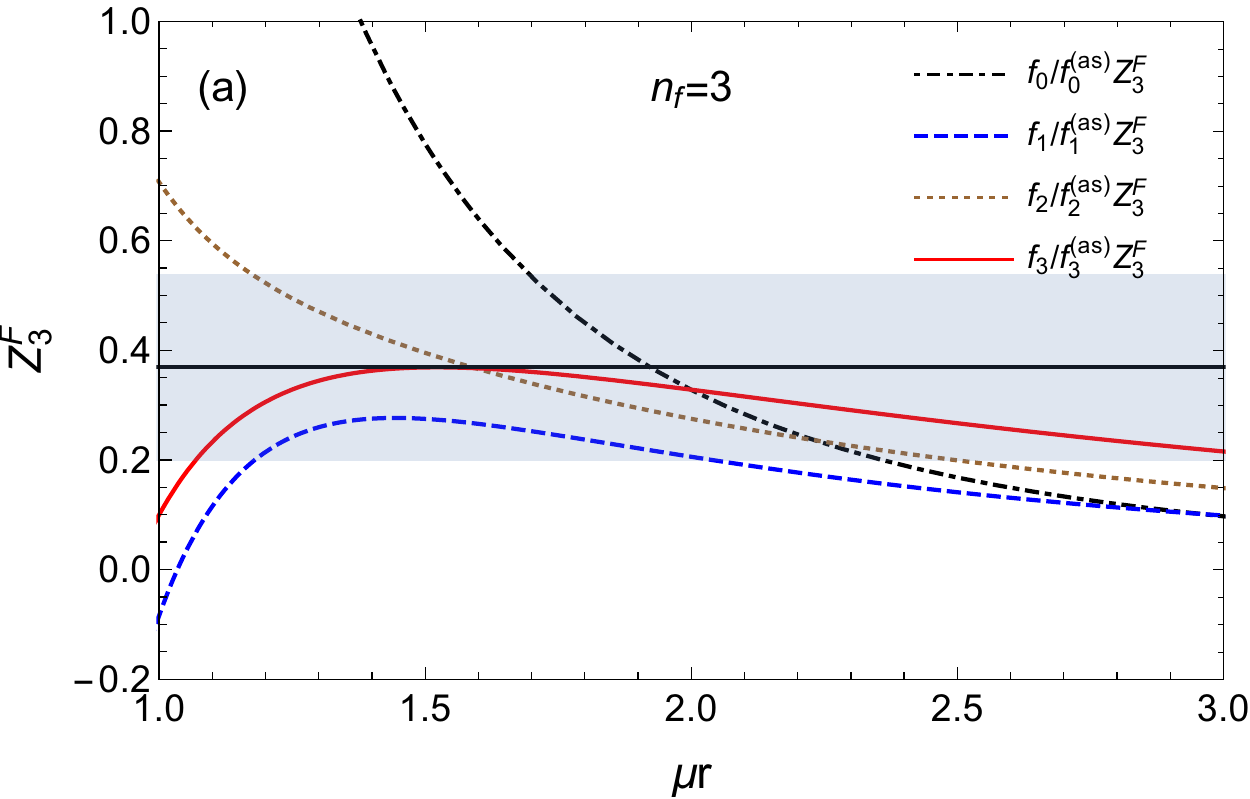}
  \end{minipage}
\begin{minipage}[b]{.49\linewidth}
  %\centering\includegraphics[width=85mm]{OldNormNf3.eps}
  \centering\includegraphics[width=80mm]{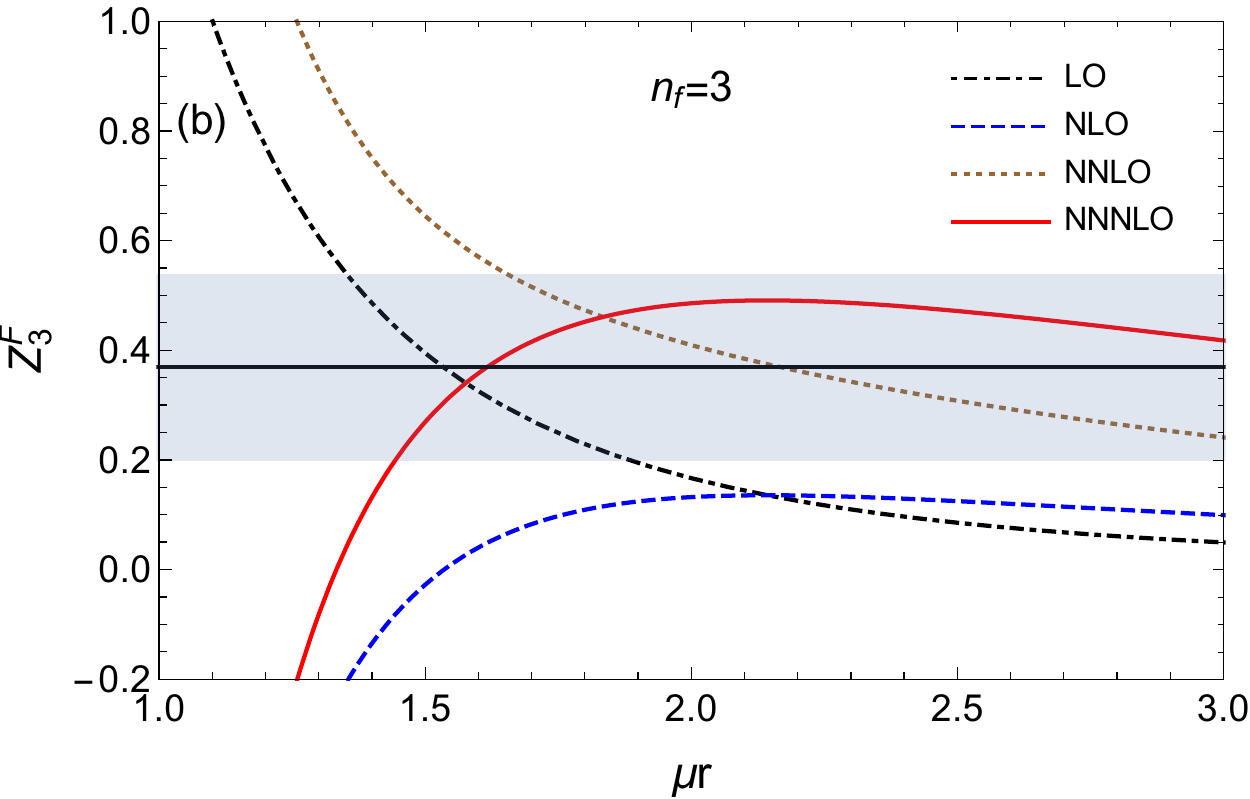}
\end{minipage}
\begin{minipage}[b]{.49\linewidth}
  %\centering\includegraphics[width=85mm]{RatioForceCoeffNf0.eps}
  \centering\includegraphics[width=80mm]{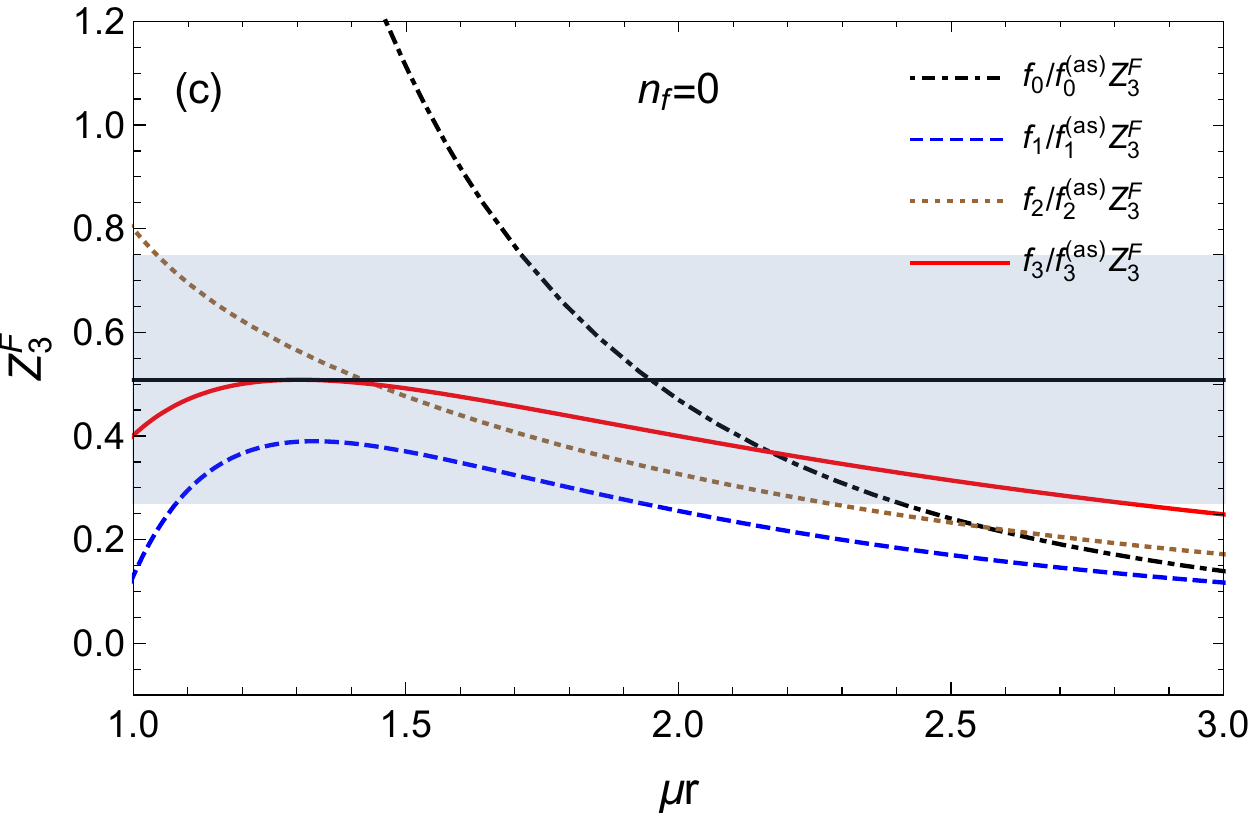}
  \end{minipage}
\begin{minipage}[b]{.49\linewidth}
  %\centering\includegraphics[width=85mm]{OldNormNf0.eps}
  \centering\includegraphics[width=80mm]{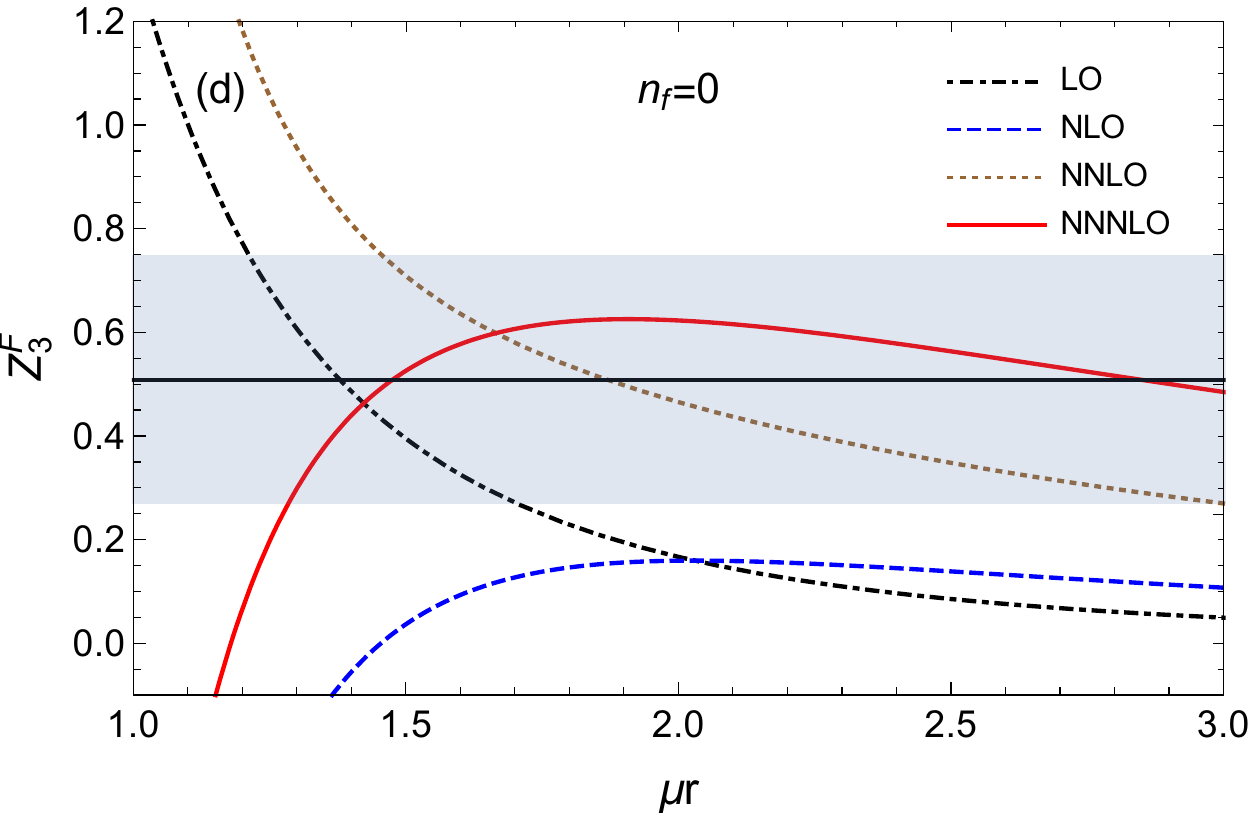}
\end{minipage}
%\vspace{-0.2cm}
\caption{\it {\bf Left figures}. Determination of $Z^F_3$ using $f_n/f_n^{(as)}Z^F_3$ as a function of $\mu r$ and for different values of $n$ in the $\MS$ scheme. {\bf Right figures}. Determination of $Z^F_3$ using \eq{NmDm}. 
Upper figures are determinations with $n_f=0$. Lower figures are determinations with $n_f=3$.} 
\label{fig:Z3}
\end{figure}

The asymptotic behavior of the coefficients of the force read:
\begin{align}
\label{fpn}
r^2f_n &\stackrel{n\rightarrow\infty}{=} Z^F_{3}(r \mu)^3\,
\left(
\frac{\beta_0}{6\pi }\right)^{\!n}
\frac{\Gamma(n+1+3b)}{\Gamma(1+3b)}
\left\{
1+\frac{3b}{n+3b}\,b_1
+
\order\left(\frac{1}{n^2}\right)
\right\}
\,,
\end{align}
and 
\be
Z^F_{3}=2Z^V_{3}
\,.
\ee
By considering the ratio of the exact and asymptotic expression we can obtain an approximate determination of the normalization of the $u=3/2$ renormalon. We obtain
\bea
\label{ZF3nf0}
\left.Z^F_{3}\right|_{n_f=0}&=&0.51^{-0.08}_{-0.21}(\Delta x)+0.05({\rm {N}^2LO})-0.10( \mathcal{O}(1/n))+0.01(\rm {us})=0.51(24) \ ,
\\
\label{ZF3nf3}
\left.Z^F_{3}\right|_{n_f=3}&=&0.37^{-0.06}_{-0.16}(\Delta x)+0.02({\rm {N}^2LO})-0.05(\mathcal{O}(1/n))+0.005(\rm {us})=0.37(17) \ .
\eea
We determine these numbers taking the central value of the ratio of $f_3/f_3^{(\rm as)}Z_3^F$ at the scale of minimal sensitivity (see Fig. \ref{fig:Z3}), which are $x \equiv \mu r=1.30$, and $x=1.52$ for $n_f=0$ and $3$ respectively.
 For the error estimate we explore different possibilities: We vary $\mu r $ by multiplying and dividing the central value by $\sqrt{2}$. This is the first error quoted in \eqs{ZF3nf0}{ZF3nf3}. We also consider the difference between 
$f_2/f_2^{(as)}Z_3^F$ and  $f_3/f_3^{(as)}Z_3^F$ at the scale of minimal sensitivity. This is the second error quoted in \eqs{ZF3nf0}{ZF3nf3}. We also estimate the importance of subleading $1/n$ corrections by considering the difference of including the $1/n$ term or not in \eq{fpn}. This is the third error quoted in \eqs{ZF3nf0}{ZF3nf3}. Finally, we also explore the importance of the ultrasoft associated terms (as they should not affect, or little, the determination of the normalization of the renormalon). The error associated to ultrasoft effects is estimated by eliminating the last term in the second line of $a_3$ in Eq. (\ref{eq:Vr}) and the last term in $f_3$ in \eq{eq:fn}. The variation is indeed small, as we show in the last error item in \eqs{ZF3nf0}{ZF3nf3}. The first and second error (and to some extent the third) are somewhat redundant, as they both measure the fact that $n=3$ is still finite. Still, for the total error, we combine all them in quadrature and make the variation symmetric around the central value. This indeed yields a conservative estimate of the error, as we can see in Fig. \ref{fig:Z3}. In Figs. \ref{fig:Z3}.(a) and \ref{fig:Z3}.(c), we can see the dependence of $Z_3^F$, i.e. of $f_n/f_n^{(\rm as)}Z_3^F$, with respect $\mu r$ for different values of $n$. Around the scale of minimal sensitivity they are inside the error band, even for a coefficient as low as 
$f_1/f_1^{(\rm as)}Z_3^F$. We profit to give determinations for other values of $n_f$ using the same error analysis. They can be found in Table \ref{Tab:Z3F}, where we also give estimates of the higher order coefficients of the perturbative series  of the force. 
\begin{table}
\centering
\label{Tab:Z3F}
\caption{Normalization constant, $Z_3^F$, of the leading renormalon of the force for different number of flavours $n_f$ in the $\MS$ scheme.}
\begin{tabular}{|c|c|c|c|c|c|c|c|}
\hline
 $n_f$ & 0 & 1 & 2 & 3 & 4 & 5 & 6 \\  \hline
 $Z_3^F$ & 0.51(24) & 0.47(22) & 0.42(20) & 0.37(17) & 0.31(14) & 0.23(10) & 0.15(8) \\ \hline
 $r^2 f_4^{(as)}$ & 8(4) & 6(3) & 4(2) & 3(1) & 1.5(7) & 0.8(3) & 0.3(2) \\ \hline
 $r^2 f_5^{(as)}$ & 31(15) & 21(10) & 13(6) & 8(4) & 4(2) & 1.9(8) & 0.65(33) \\ \hline
\end{tabular}
\end{table}

Our conclusions are different from those in Ref. \cite{Sumino:2020mxk}, where it was concluded that it was not possible to determine the normalization of the $u=3/2$ renormalon. The authors used the function 
\begin{align}
D^{(N)}_F(u)&=\sum_{n=0}^{N}D_F^{(n)} u^n=(1-\frac{2}{3}u)^{1+b}B^{(N)}[F](t(u))
\\
\nn
&
=Z_3^F\frac{1}{r^2}\left(1+c_1(1-\frac{2}{3}u)+c_2(1-\frac{2}{3}u)^2+\cdots
\right)+(1-\frac{2}{3}u)^{1+b}({\rm analytic\; term})\,,
\end{align}
following the method proposed in Refs.~\cite{Lee:1996yk,Lee:1999ws}, and first quantitatively applied to the leading renormalon of the pole mass and static potential in Ref.~\cite{Pineda:2001zq}.
$D^{(N)}_F(u)$ is singular but bounded at the first IR renormalon.
Therefore, we can estimate $Z_3^F$ from the first
coefficients of the series in $u$, using 
\be
\label{NmDm}
Z_3^F\frac{1}{r^2}=D^{(N)}_F(u=3/2)\,.
\ee
We plot the predictions for different orders $N$ in the figures~\ref{fig:Z3}.(b) and ~\ref{fig:Z3}.(d) for $n_f=3$ and $n_f=0$ respectively. We observe that the convergence is worse than for the determination of $Z^F_3$ using $f_n/f_n^{(as)}Z^F_3$. This was also observed very clearly in \cite{Bali:2013pla} for the energy of an static source. In that case, and this case here, we observe convergence but at a slower pace. Actually, the NNLO and NNNLO predictions are well inside the error band of our predictions in Eqs.~\eqref{ZF3nf0} and \eqref{ZF3nf3}, though
less precise, as the variation between different orders is bigger than in the previous case. Compared with the analysis in \cite{Sumino:2020mxk}, we make the analysis at larger values of $x$, but close to one, where we find stability. For the method using \eq{NmDm} stability is found for $x \sim 2$. For this method, working with $x=1$ does not yield a convergent series. This may explain the conclusions reached in \cite{Sumino:2020mxk}.

\section{Fit of $\al$}
\label{Sec:Fit}
We will now compare our theoretical expressions for the static energy with recent lattice data obtained with $n_f=3$ active flavours. The static energy computed in the lattice can be equated with the theoretical expressions we have up to a constant. Therefore, we will always use the following equality:
\be
\label{Edifference}
E^{\rm latt}(r)-E^{\rm latt}(r_{ref})=E^{\rm th}(r)-E^{\rm th}(r_{ref})
\,.
\ee
In principle the analysis should not depend on the value of $r_{ref}$ we use in this equation. In practice there will be some dependence (we will check this dependence later). By default we will take the value $r_{ref}=r_{\rm min}$, i.e. the point at the shortest distances we use.

$E^{\rm th}(r)-E^{\rm th}(r_{ref})$ is renormalon free. Nevertheless, there are different ways to implement this cancellation which are not equally efficient. The use of ${\cal F}(r)$ seems optimal in this respect. On the one hand the leading renormalon identically vanishes. The subleading renormalons of the static potential cancel with those of the ultrasoft energy. This cancellation takes place order by order in $\alpha$ if both quantities are expanded in powers of $\alpha$ evaluated at the same scale, and if both perturbative expansions are truncated at the same order. This is something that we will not do, as the perturbative expansion of the static potential and the ultrasoft energy are known to different orders, and the natural energy scales in $F_{\rm PV}$ and $\frac{d}{dr}\delta E_{us}^{\rm PV}(r;\nu_{us})$ are different. This last issue also reflects in that setting $\nu_s=\nu_{us}$ misses the resummation of large logarithms associated to the ultrasoft scale. These can be important, and not incorporating them can jeopardize the convergence of the perturbative series. Therefore, we also perform the resummation of logarithms, and our default expression will be the RG improved expression. In this case the precision we have is LL, NLL, NNLL and NNNLL if we do not include renormalon effects. From the analysis performed in Sec. \ref{u}, we have seen that perturbation theory of the force has reached high enough orders to be sensitive to the $d=3$ renormalon. Thus, to the NNNLL expression, we will add the terminants associated to the $d=3$ renormalon. We will name this approximation NNNLL$_{\rm hyp}$. In the hyperasymptotic counting, this means that the maximal accuracy that we will seek in ${\cal F}$ will be $(3,0)$. We will take as default that the asymptotic behavior associated to the $d=3$ renormalon is reached for $N=3$ for the force. For the ultrasoft term we will then take $N=0$, since we will only incorporate one term of the perturbative expansion of the ultrasoft term. The different order at which we truncate the two perturbative expansions, and the different scale they depend on, make the terminants associated to $F(r)$ and $\frac{d}{dr}\delta E_{us}$ to be different. Obviously, the difference of determinations using NNNLL or NNNLL$_{\rm hyp}$ will allow us to see the impact of including the terminants\footnote{Alternatively, we have also performed fits changing the order at which we start including the terminant  in the static potential from three to two. We indeed find the variation to be small.  For a fixed scale $\nu$, the change of the order at which we start including the terminant is implemented by changing the value of $c_F$. We then find that our fits have an small dependence on $c_F$.}. 

In order to compare with analyses where the resummation of logarithms is not incorporated, we will also perform computations with $\nu_s=\nu_{us}$. In this case, if we neglect renormalons, we can compute the observable with LO, NLO, NNLO and NNNLO accuracy (note that LO=LL and NLO=NLL), accordingly to the order in $\alpha$ we truncate the perturbative series of ${\cal F}(r)$. To make the connection smooth with the RG improved expression, we also incorporate at NNNLO the terminants associated to the $u=3/2$ renormalons of $F$ and $\frac{d}{dr}\delta E_{us}$. Note that for this last term we will still use $N=0$, even though the renormalization scale of $\alpha$ is bigger. We will discuss this issue later. 

To compare with the lattice results, we have to integrate over $r$:
\be
\label{Eforceth}
E^{\rm th}(r)-E^{\rm th}(r_{ref})=\int_{r_{ref}}^r dr' {\cal F}^{\rm RG}(r')
\,,
\ee
and to fit the outcome with the lattice data to determine $\lQ$. As we have mentioned above, our default fit is made using the RG improved expressions. For the central values of the renormalization scales, we take $(\nu_s,\nu_{us})=(1/r,C_A\al(\nu_s)/(2r))$. In the following we perform such fit and quantify the different sources of error. Note that ${\cal F}^{\rm RG}(r')$ has to be introduced in the integral order by order in $\alpha$ in order to implement the renormalon cancellation in \eq{Eforceth}, as it was first explained in \cite{Pineda:2002se}. \eq{Eforceth} was originally used in \cite{Necco:2001gh}, but its use for competitive determinations of $\Lambda_{\MS}$ was first made in \cite{Bazavov:2014soa}. 

{\bf Dependence on the data points}\\
For the fits, we use the lattice data of \cite{Bazavov:2019qoo} (supplemented with the information given in \cite{Bazavov:2018wmo}), which has made an updated error analysis of the data of \cite{Bazavov:2017dsy}. Of these data points we only consider those obtained with $\beta=8.4$, as they correspond to the shortest distances available: $1/a \simeq 8.3$ GeV. In this ensemble the strange quark mass has been fine tuned to its physical value, and the pion mass gets the value 320 MeV in the continuum. This is only statistically significant\footnote{We thank J.H. Weber for informing us of this.} for $r >0.4 r_1 \sim 1/1.6$ GeV$^{-1}$ (see \cite{Weber:2018bam}). In the fits, we will approximate the light quark masses to zero. The uncertainty associated with fixing the physical units of the parameter $r_1$ was seen in \cite{Bazavov:2019qoo} to be comparatively small compared with other uncertainties. Therefore, we will neglect it in the following. It was also observed in this reference that the effect of the correlation of the points to the final error was small. Thus, we also neglect this source of error. The discretization errors depend on the size of the parameter $r/a$. They have been studied in detail in \cite{Bazavov:2019qoo}, where it was concluded that, for $r/a\leq \sqrt{8}$, tree-level improvement was enough to bring the discretization errors down to the point that they were smaller than the statistical errors and could, in comparison, be neglected. Therefore, we will use tree-level improved data and disregard the lattice data at shortest distances (for $r/a\leq \sqrt{8}$), as well as the special geometry $r/a=\sqrt{12}$. This corresponds to one of the methods followed in \cite{Bazavov:2019qoo} to account for discretization effects. This means that the shortest distance we consider is $r_{\rm min}=2.827 \, a$, which in physical units reads $r_{\rm min}=0.353$ GeV$^{-1}$. We have also compared with the older unquenched data of \cite{Cheng:2007jq}. Overall, we observe the same qualitative features, though the lattice errors are bigger. Therefore, we will only present quantitative analyses with the data of \cite{Bazavov:2019qoo}. 

To test the sensitive of the fit to the data we consider different ranges of data (similarly as it was done in \cite{Bazavov:2014soa}). 
We consider the following ranges: Set I: 0.353 GeV$^{-1}\leq r \leq 0.499$ GeV$^{-1}$, Set II: $0.353$ GeV$^{-1}\leq r \leq 0.612$  GeV$^{-1}$, 
Set III: $0.353$ GeV$^{-1}\leq r \leq 0.8002$ GeV$^{-1}$ and Set IV: $0.353$ GeV$^{-1} \leq r \leq 1$ GeV$^{-1}$. 
The number of data points of each set is 8, 17, 31 and 50, respectively. 

\begin{figure}[htb] 
%\unitlength=1mm
\centering\includegraphics[width=140mm]{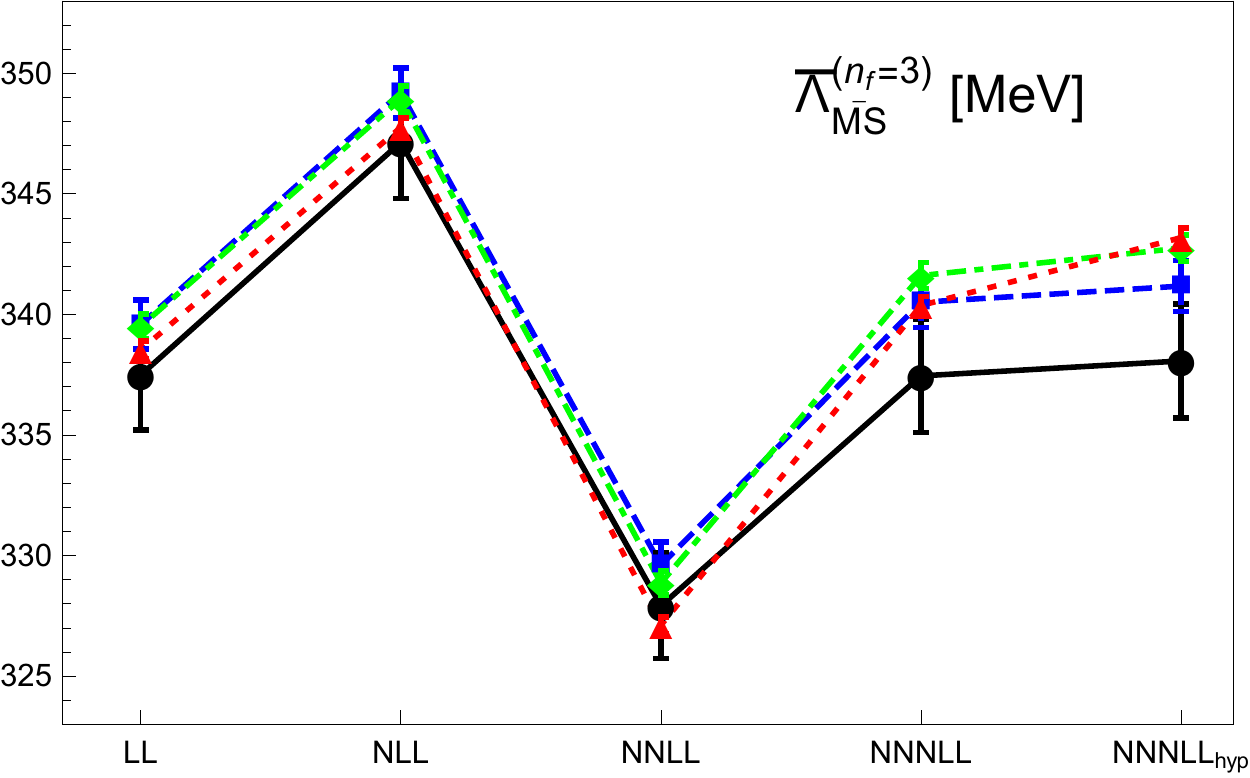}
   \centering\includegraphics[width=140mm]{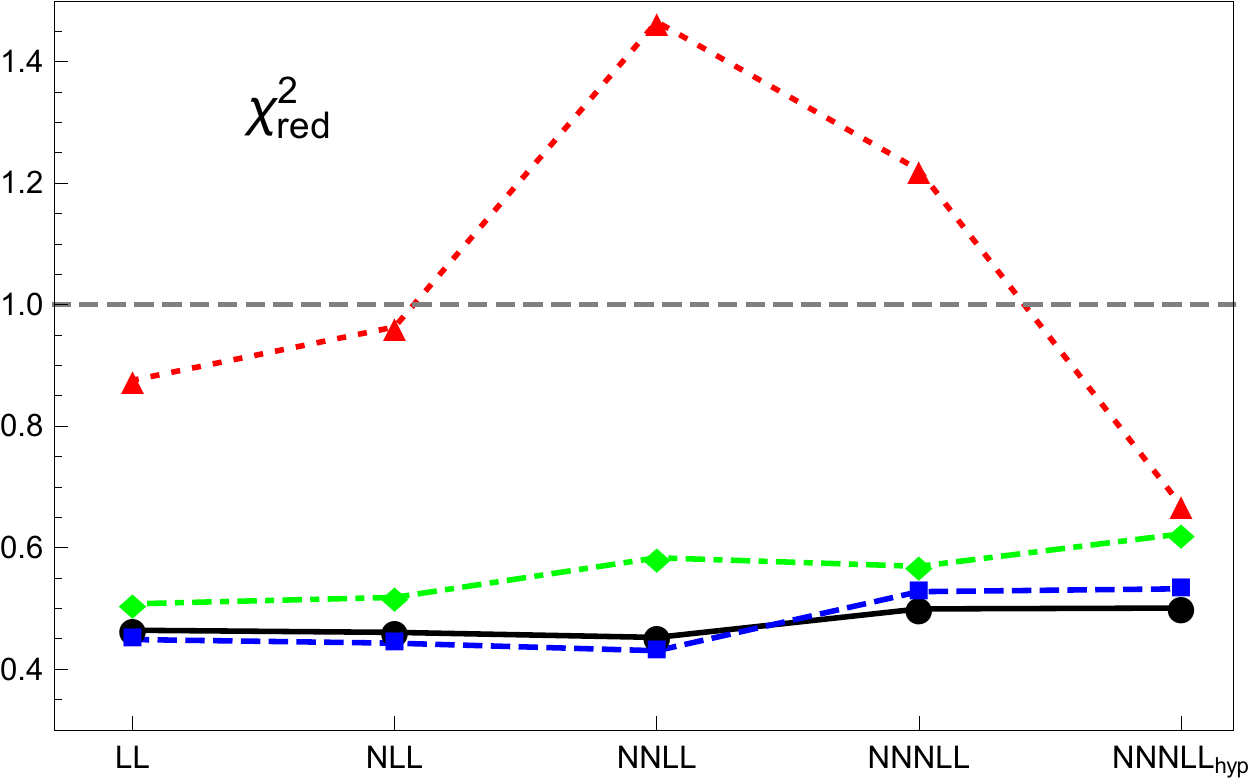}
%  \centering\includegraphics[width=140mm]{FitDatarref.pdf}
%  \centering\includegraphics[width=140mm]{FitDataChi2rref.pdf}
%\vspace{-0.2cm}
\caption{{\bf Upper panel} Determination of $\Lambda_{\MS}^{(n_f=3)}$ at LL, NLL, NNLL, NNNLL and NNNLL$_{hyp}$ using the data sets: Set I (continuous black line), Set II (dashed blue line), Set III (dash-dotted green line), and Set IV (dotted red line). The error displayed is only the statistical error of the fits.
{\bf Lower panel} $\chi^2_{\rm red}$ for the fit of $\Lambda_{\MS}^{(n_f=3)}$ at LL, NLL, NNLL, NNNLL and NNNLL$_{hyp}$ using the data sets (continuous black line), Set II (dashed blue line), Set III (dash-dotted green line), and Set IV (dotted red line).} 
\label{fig:FitData}
\end{figure}

We show the result in Fig. \ref{fig:FitData}. We observe the following. The dependence of the value of $\Lambda_{\MS}$ on the range of the data set is very small. Obviously, as we increase the number of points, the statistical errors get smaller. This small dependence holds irrespectively of the order in the approximation for the theoretical expression used. We only see very small differences at NNNLL and NNNLL$_{hyp}$ order between the value obtained from the data set I and the rest (within one sigma for the statistical error, which is the only one we display in Fig. \ref{fig:FitData}), and basically vanishing between the data sets II, III and IV.

To see how reliable the results are, we study the reduced $\chi^2$ obtained with each data set. For the data sets I and II, the fit yields $\chi^2_{\rm red} \sim 0.5$ to all orders in the hyperasymptotic expansion. Therefore, there is no significant dependence on the number of data points. For the data Set III, there is a mild increase: $\chi^2_{\rm red} \sim 0.5-0.6$ but still well below 1. It is when we consider data set IV, which includes points down to $1/r=1$ GeV, that we see a significant increase in the $\chi^2_{\rm red}$. The magnitude of this increase, however, depends on the order, and, even in the worst case, it is not much bigger than 1. The  LL and NLL fits yield  $\chi^2_{\rm red}$ slightly below 1, with a slight increase when going from LL to NLL. The $\chi^2_{\rm red}$ reaches the maximum, 1.46, at NNLL. Since then higher order fits improve the quality of the fit and, significantly, the NNNLL$_{hyp}$ fit, which adds the terminant of the $u=3/2$ renormalon, yields a $\chi^2_{\rm red} =0.67$, similar to the other data sets. The Set IV is the more sensitive to the infrared, as it goes down to $1/r \sim 1$ GeV. This may reflect in a larger sensitivity to ultrasoft associated physics, which will then need to be described more accurately. This matches with what we see for the $\chi^2_{\rm red}$ with set IV: LL, NLL, NNLL and NNNLL show a bigger $\chi^2_{\rm red}$ (we emphasize, though, that they are still of order 1), which then goes down to a value similar to the one obtained with the other data sets after the inclusion of the terminants.

This result is not trivial. We expect more sensitivity to infrared physics with the data set IV. What is not trivial is that this larger sensitivity to the infrared can be well described by our weak-coupling analysis. Indeed it is surprising that the ultrasoft effects do not blow up in any of the fits, since the $\alpha(\nu_{us})$ is evaluated at a rather low scale. For illustration (to produce these numbers we take $\Lambda_{\MS}^{(n_f=3)}=330$ MeV), for Set I $\alpha(\nu_{us}) \in (0.46, 0.57)$,  for Set II $\alpha(\nu_{us}) \in (0.46, 0.65)$, for Set III $\alpha(\nu_{us}) \in (0.46, 0.75)$, and for Set IV $\alpha(\nu_{us}) \in (0.46,0.78)$. For this last data set, the very last points with smaller energy reach a regime where $\al(1/r)$ grows faster than $1/r$, so that $\nu_{us}$ grows for them. Therefore, their inclusion in fits should be taken with caution. 

Overall, by only looking at the $\chi^2_{\rm red}$, we do not have a clear signal of which data set to use and, indeed, the fits yield similar numbers and $\chi^2_{\rm red}$ at NNNLL$_{hyp}$. Therefore, we will use set I as it is less sensitive, in principle, to long distances, though, as we said, the $\chi^2_{\rm red}$ of the fits does not give a clear signal of a deterioration of the quality of the fit (something that one would expect if our perturbative approximation were not a good approximation to the data). In this respect note that the data set IV, which is more sensitive to the ultrasoft scale, yields a good $\chi^2_{\rm red}$ after the introduction of the terminants associated to the $u=3/2$ renormalon. We interpret this as an indication that the ultrasoft effects can be well described by a weak-coupling computation even at scales as low as $\nu_s \sim 1$ GeV. 

Another motivation to use the data set I is that the $\beta=8.4$ ensemble suffers from frozen topological charge in the Montecarlo evolution. It has been shown (see \cite{Weber:2018bam}) that the effects of frozen topology in different sectors are statistically irrelevant  for $r<0.4 r_1\sim 1/1.61$ GeV$^{-1}$. Therefore, by using the data set I this problem is completely avoided. On top of that, as mentioned above, the effects due to finite light-quark masses are not statistically significant for this energy range.

\begin{figure}[htb] 
%\unitlength=1mm
\centering\includegraphics[width=140mm]{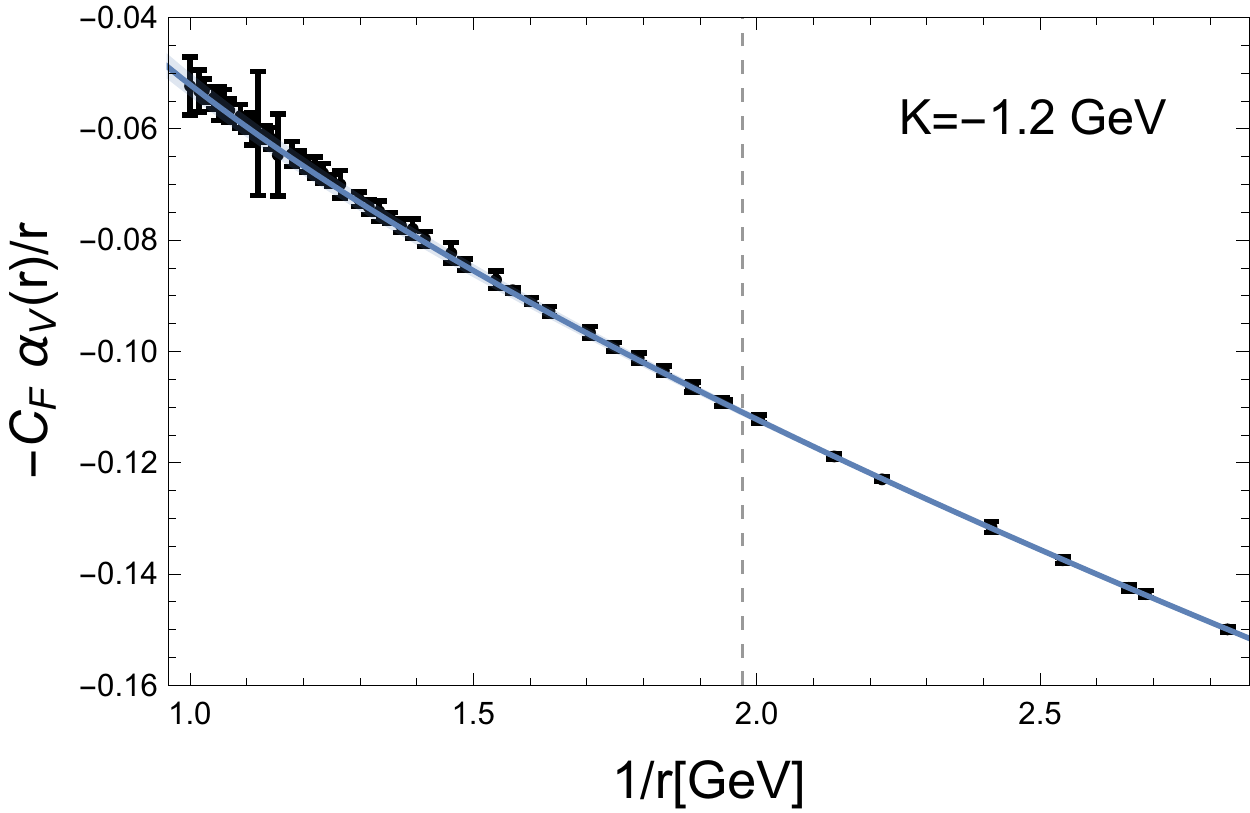}
   \centering\includegraphics[width=140mm]{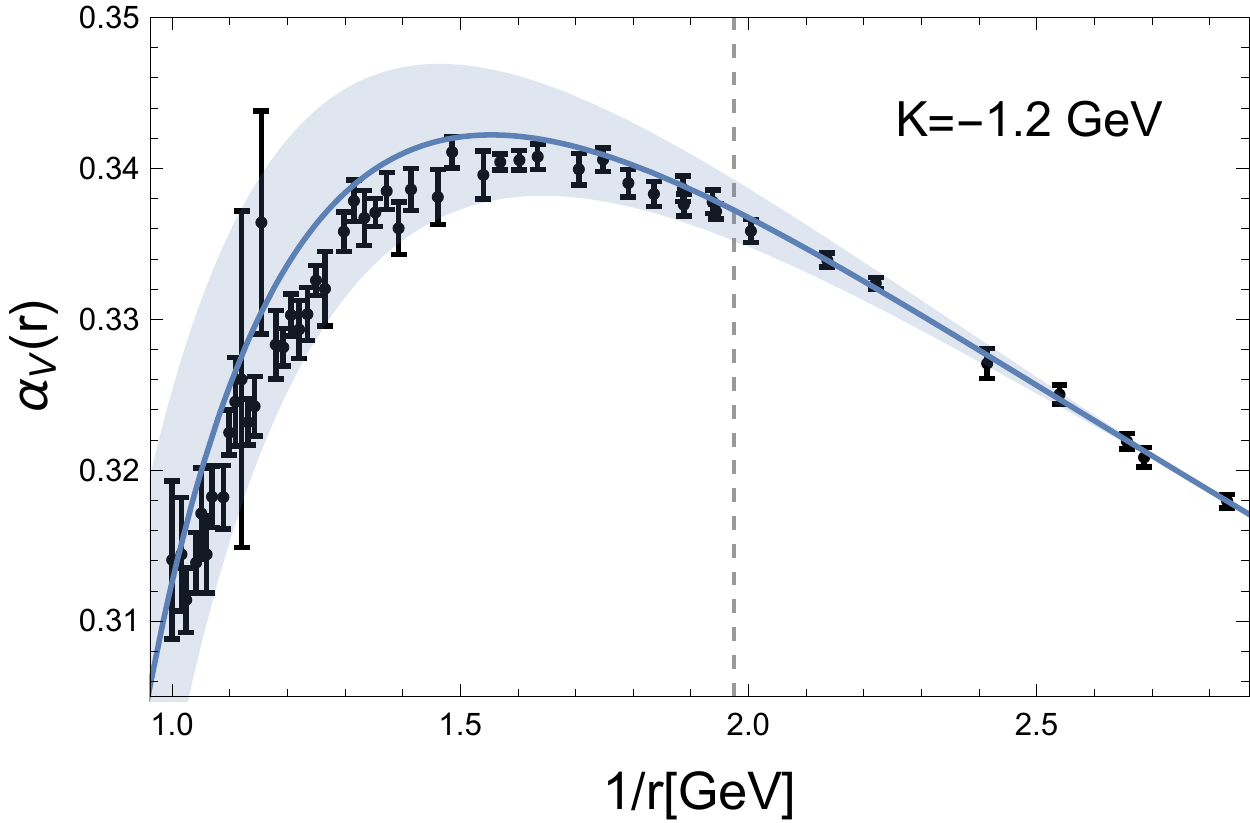}
%\centering\includegraphics[width=120mm]{figrVext.pdf}
%\centering\includegraphics[width=120mm]{PreliminaryPlot.png}
%\vspace{-0.2cm}
\caption{{\bf Upper panel}: $-C_F\frac{\alpha_V(r)}{r}\equiv \int_{r_{ref}}^r dr' {\cal F}^{\rm RG}(r')+K$ at NNNLL$_{hyp}$ with $\Lambda_{\MS}^{(n_f=3)}=338(12)$ MeV and $K=-1.2$ GeV (solid blue line and blue band) versus the lattice points $E^{\rm latt}(r)-E^{\rm latt}(r_{ref})+K$. Only points to the right of the vertical dashed line are included in the fit.
{\bf Lower panel}:  $\alpha_V(r)$ at NNNLL$_{hyp}$ with $\Lambda_{\MS}^{(n_f=3)}=338(12)$ MeV and $K=-1.2$ GeV (solid blue line and blue band) versus the lattice points $ -\frac{r}{C_F}\left(E^{\rm latt}(r)-E^{\rm latt}(r_{ref})+K\right)$. Only points to the right of the vertical dashed line are included in the fit.} 
\label{fig:comparisonPot}
\end{figure}

In order to see the quality of our fit, we also compare our theoretical expression using the values of $\Lambda_{\MS}^{(n_f=3)}$ obtained from the fit with the lattice data. It is customary to compare directly with the potential (this can be done after fixing a normalization constant $K$ that we fix below). The comparison is very good in the whole range we compare (up to 1 GeV), as we can see in the upper panel of Fig. \ref{fig:comparisonPot} (note that we plot as a function of $1/r$, a plot in terms of $r$, as it is customarily done, is even less precise). Nevertheless, such comparisons do not allow us to see the fine details due to the dependence in powers of $r$ of the potential. For such comparison, it is better to define 
\be
\alpha_V(r)\equiv -\frac{r}{C_F}\left(
\int_{r_{ref}}^r dr' {\cal F}^{\rm RG}(r')+K\right)
\,,
\ee
and we adjust $K$ such that most of the $r$ dependence vanishes. We show the comparison in the lower panel of Fig. \ref{fig:comparisonPot}. 
It is remarkable that pure perturbation theory predicts very well the data down to 1 GeV. The error band perfectly encodes all the data. This means, in particular, that with the precision of our computation we do not see any trace of nonperturbative effects down to scales $1/r \sim$ 1 GeV.

\begin{figure}[htb] 
%\unitlength=1mm
   \centering\includegraphics[width=140mm]{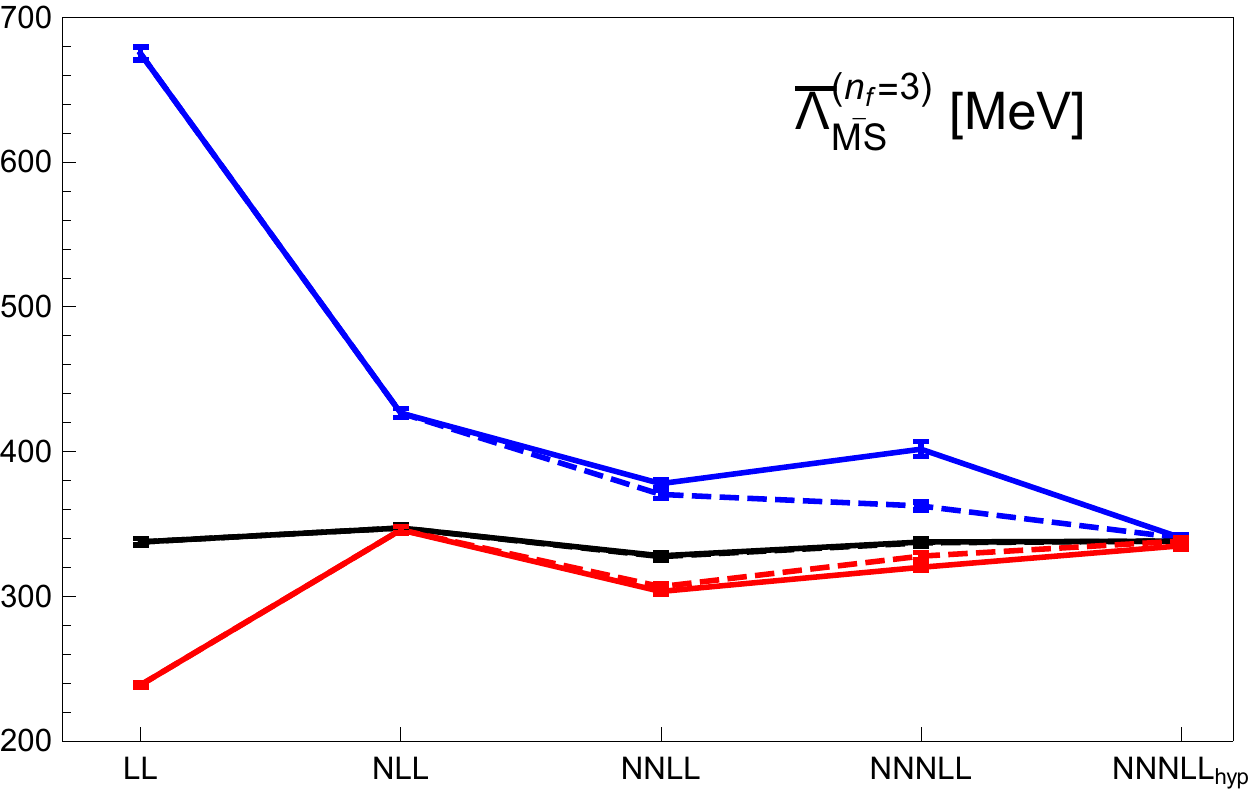}
%\vspace{-0.2cm}
\caption{ Determination of $\Lambda_{\MS}^{(n_f=3)}$ at LL, NLL, NNLL, NNNLL and NNNLL$_{hyp}$ using the data set I with $\nu_{us}=C_A\al(\nu_s)/(2r)$ and $\nu_s=2/r$ (continuous blue line), $\nu_s=1/r$ (continuous black line) and $\nu_s=1/(\sqrt{2}r)$ (continuous red line). We also plot the determination of $\Lambda_{\MS}^{(n_f=3)}$ at LL, NLL, NNLL, NNNLL and NNNLL$_{hyp}$ using the data set I with $\nu_{us}=1$ GeV and $\nu_s=2/r$ (dashed blue line), $\nu_s=1/r$ (dashed black line) and $\nu_s=1/(\sqrt{2}r)$ (dashed red line). With this scale resolution the continuous and dashed black lines are hardly distinguishable. This also happens to a large extent with the continuous and dashed red lines. The error displayed is only the statistical error of the fits.} 
\label{fig:FitData05}
\end{figure}

{\bf Dependence on $\nu_s$}\\
We now test the sensitivity of the fit on $\nu_s$. We will mainly work with the data set I, with which we can do variations of the parameters without entering in the regime where perturbation theory breaks down. We will try to vary $\nu_s$ but keeping $\nu_{us}$ constant. Our central value for $\nu_{us}$ is $\nu_{us}=C_A\al(\nu_s)/(2r)$. For the data Set I this yields values around $\nu_{us}=1$ GeV. Therefore, besides $\nu_{us}=C_A\al(\nu_s)/(2r)$, $\nu_{us}=1$ GeV  will be the other choice we take for $\nu_{us}$. 
We observe that both fits, with $(\nu_s,\nu_{us})=(1/r,C_A\al(\nu_s)/(2r))$ and with $(\nu_s,\nu_{us})=(1/r,1$ GeV) yield very similar results. We show the outcome in Fig. \ref{fig:FitData05}. Indeed, in the figure, the fits with $(\nu_s,\nu_{us})=(1/r,C_A\al(\nu_s)/(2r))$ (continuous black line) and with $(\nu_s,\nu_{us})=(1/r,1$ GeV)  (dashed black line) are hardly distinguishable with the resolution set by the figure. This is so to all orders. They also kind of oscillate around the central value. Actually, at the NNNLL$_{hyp}$ level, the difference between fits is less than 0.1 MeV !!

For the variation of $\nu_s$ we take $\nu_s=x_s/r$ within the range $x_s \in [1/\sqrt{2},2]$. The lower limit of $\nu_s$ is chosen to avoid reaching scales too low for our weak coupling analysis to break down. This happens, both if we take $(\nu_s,\nu_{us})=(1/(2r),1$ GeV) or  $(\nu_s,\nu_{us})=(1/(2r),C_A\al(\nu_s)/(2r))$. As we increase $\nu_s$, things significantly improve, and a safe value to take as a lower limit is $x_s=1/\sqrt{2}$. We show the results of the fits for the central and extreme values of the parameter in Fig. \ref{fig:FitData05}. 

We now compare with the fits with $\nu_s=2/r$. It is interesting to see the different behavior of the perturbative expansion if we work with $\nu_s=1/r$ or with $\nu_s=2/r$. 
Somewhat, working with $\nu_s=1/r$ gives the right result from the beginning, and adding more terms of the perturbative expansion makes the result to oscillate around the central value. On the other hand, working with $\nu_s=2/r$, the LL result is quite off the expected result, but then adding higher order terms of the perturbative expansion makes the prediction to converge to the same result we have obtained with $\nu_s=1/r$. The convergence is perfect within the statistical errors, and also irrespectively of working with $\nu_{us}=C_A\al(\nu_s)/(2r)$ or with $\nu_{us}=1$ GeV. Still, the convergence pattern is not equal in these two cases for $\nu_s=2/r$. Not taking an optimal $\nu_s$ ($\sim 1/r$) makes the determinations with $(\nu_s,\nu_{us})=(2/r,C_A\al(\nu_s)/(2r))$ or with $(\nu_s,\nu_{us})=(2/r,1$ GeV) to be significantly different at NNNLL. Nevertheless, this difference is nicely elliminated after the inclusion of the  terminants associated to the $u=3/2$ renormalon. Even more so, the inclusion of the terminants associated to the $u=3/2$ renormalon is also fundamental to get agreement of these fits with the fits with $\nu_s=1/r$. A similar discussion holds for the case with $\nu_s=1/(\sqrt{2}r)$, though the overall behavior is better: The LL result is closer to the value obtained with $\nu_s=1/r$, and the difference between the determinations with $(\nu_s,\nu_{us})=(1/(\sqrt{2}r),C_A\al(\nu_s)/(2r))$ or with $(\nu_s,\nu_{us})=(1/(\sqrt{2}r),1$ GeV) at NNNLL is small. Finally, for $(\nu_s,\nu_{us})=(1/(\sqrt{2}r),1$ GeV) and 
$(\nu_s,\nu_{us})=(1/(\sqrt{2}r),C_A\al(\nu_s)/(2r))$, we get $\Lambda_{\MS}^{(n_f=3)}=338$ MeV and $\Lambda_{\MS}^{(n_f=3)}=335$ MeV  respectively with the NNNLL$_{hyp}$ theoretical expression. We then conclude that in the range $\nu_s \in (1/(\sqrt{2}r,2/r)$ the result is stable at the 2 MeV level if we fix $\nu_{us}=$1 GeV. The spread of the result is slightly larger, at around the 3 MeV level, if we set $\nu_{us}=C_A\al(\nu_s)/(2r)$ instead. This very tiny increase can be interpreted by the fact that $\nu_{us}$ is not completely constant as we change $x_s$, since $\nu_{us}=C_A\al(x_s/r)/(2r)$ is not exactly equal to $\nu_{us}=C_A\al(1/r)/(2r)$, the value we use in our reference fit. In the next item, we will study the dependence of our fits with more extensive variations of $\nu_{us}$.  For $\nu_s$, we conclude that, with the present level of precision reached by the theoretical expression, the dependence on $\nu_s$ of the fit, of order $\sim 2$ MeV, can be neglected compared with other uncertainties.

In the whole parameter range we have studied, the $\chi^2_{red}$ is reasonable. Therefore, all fits are equally good in this respect. The only exception is the NNNLL prediction for 
$(\nu_s,\nu_{us})=(2/r,C_A\al(\nu_s)/(2r))$, which has a $\chi^2_{red}\simeq 1.9$. We find then significant that it moves away from the convergent pattern that is observed in the other fits. It is also then significant that the inclusion of the terminant brings agreement with the other fits and lowers the $\chi^2_{red}$ down to $\chi^2_{red}=0.42$, much below 1. 

{\bf Dependence on $\nu_{us}$}\\
We now test the sensitivity of the fit on $\nu_{us}$. In order to keep the hierarchy of scales between the soft and ultrasoft scale, we have varied them in a correlated way as a function of a single parameter $x$: 
\be
\label{eq:x}
(\nu_s,\nu_{us})=\left(x\frac{1}{r},x\frac{ C_A\al(x/r)}{2r}\right)
\,.
\ee 
The range we take for $x$ is 
$x \in [1/\sqrt{2},2]$, similarly as we did in the previous item. For the ultrasoft scale, we do so because, otherwise, we reach very low values for $\nu_{us}$ that make $\al(\nu_{us})$ to blow up. This happens, for instance, if we take $(\nu_s,\nu_{us})=(1/(2 r),C_A\al(\nu_s)/(4 r))$. In this case, for the data Set I, the fit significantly deteriorates with a $\chi^2_{\rm red}\simeq 4$ and $\Lambda_{\MS}^{(n_f=3)}\simeq 411$ MeV. As we increase $\nu_s$, things significantly improve, and for $(\nu_s,\nu_{us})=(1/(\sqrt{2}r),C_A\al(\nu_s)/(2\sqrt{2}r))$, we get $\Lambda_{\MS}^{(n_f=3)}=347$ MeV  with $\chi^2_{\rm red}=0.56$. This yields a difference of 9 MeV with the central value $\Lambda_{\MS}^{(n_f=3)}=338$ MeV obtained with $(\nu_s,\nu_{us})=(1/r,C_A\al(\nu_s)/(2r))$. This is the same difference (with the same sign) as obtained with $(\nu_s,\nu_{us})=(2/r,2C_A\al(\nu_s)/(2r))$. In this respect the fit with $(\nu_s,\nu_{us})=(1/r,C_A\al(\nu_s)/(2r))$ can be considered a (close to the) minimum within the families of fits with $(\nu_s,\nu_{us})=\left(x\frac{1}{r},x\frac{ C_A\al(x/r)}{2r}\right)$. 

\begin{figure}[htb] 
%\unitlength=1mm
  \centering\includegraphics[width=150mm]{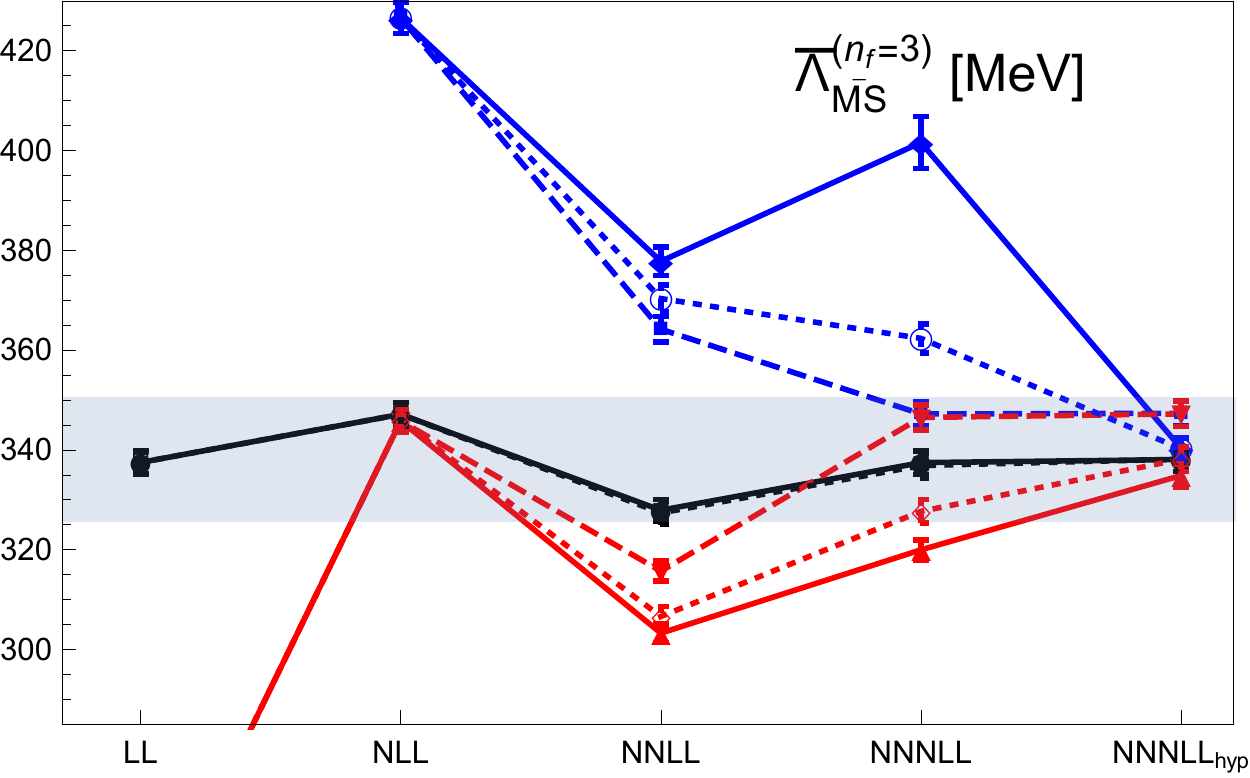}
%\vspace{-0.2cm}
\caption{Determination of $\Lambda_{\MS}^{(n_f=3)}$ at LL, NLL, NNLL, NNNLL and NNNLL$_{hyp}$ using the data set I 
with different options for the (soft, ultrasoft) scale: 
{\bf A)} 
with $(\nu_s, \nu_{us})=(1/r, C_A\al(\nu_s)/(2r))$ (continuous black line with filled black points), 
with $(\nu_s, \nu_{us})=(2/r, C_A\al(\nu_s)/(2r))$ (continuous blue line with filled blue diamonds), 
and with $(\nu_s, \nu_{us})=(1/(\sqrt{2}r),C_A\al(\nu_s)/(2r))$ (continuous red line with filled red triangles); 
{\bf B)}  
with $(\nu_s, \nu_{us})=(2/r, 2C_A\al(\nu_s)/(2r))$ (dashed blue line with filled blue squares), 
and with $(\nu_s, \nu_{us})=(1/(\sqrt{2}r),C_A\al(\nu_s)/(2\sqrt{2} r))$ (dashed red line with filled red inverted triangles); 
and 
{\bf C)}
with $(\nu_s, \nu_{us})=(1/r, 1$ GeV) (dotted black line with empty black points), 
with $(\nu_s, \nu_{us})=(2/r, 1$ GeV) (dotted blue line with empty blue points), 
and with $(\nu_s, \nu_{us})=(1/(\sqrt{2}r),1$ GeV) (dotted red line with empty red diamonds). 
The error displayed is only the statistical error of the fits. We also show the error band generated by our prediction \eq{eq:LambdaFinal}. Note that the resolution in this figure has been increased with respect to the one in Fig. \ref{fig:FitData05}. }
\label{fig:FitUS}
\end{figure}

We show our results in Fig. \ref{fig:FitUS}. The agreement is very good. The difference is of order 9 MeV between the $x=2$ and $x=1$ fits and also between the $x=1$ and $x=1/\sqrt{2}$ fits. The convergence is already reached at the NNNLL level. We observe that, by correlating the soft and ultrasoft scale as done in \eq{eq:x}, the convergence is already reached at the NNNLL level, and the contribution of the terminant associated to the $u=3/2$ is very small. Note that this was not so when we took
$\nu_{us}= C_A\al(x/r)/(2r)$ or $\nu_{us}=1$ GeV. For illustration we show again the fits with $\nu_{us}= C_A\al(x/r)/(2r)$ and with $\nu_{us}=1$ GeV in Fig. \ref{fig:FitUS}.
In that case the terminant contribution is crucial to get agreement between fits with different values of $x_s$. This reflects that the terminant plays a crucial role to diminish the dependence in $\nu_{us}$ and to get convergence to the same value irrespectively of how we correlate the soft with the ultrasoft scale.  We take the largest difference between the different possibilities we have considered ($\sim 9$ MeV) as an estimate of higher order effects of perturbation theory. Notice that the cancellation of the ultrasoft scale dependence comes from several places. On the one hand we have the perturbative contribution of $\frac{d}{dr}\delta E_{us}$ (the magnitude of this contribution is small), we have the contribution from the derivative of $\delta V_{RG}$, and finally the contribution from the terminant associated to the perturbative series of $\frac{d}{dr}\delta E_{us}$. Note that we only include one term of the pertubative expansion in powers of $\al(\nu_{us})$ in $\frac{d}{dr}\delta E_{us}$. One may be worried then that $N_P=0$ is too low for the incorporation of the terminant associated to the ultrasoft energy. Nevertheless, we are working at rather low scales. We will indeed see in \Sec{Sec:FitV} for the direct comparison of the static potential that the asymptotic behavior can easily set in at basically the lowest order.   

{\bf Dependence on $r_{ref}$}\\
The fit should be independent of $r_{ref}$. In practice, however, the result may depend on the value of $r_{ref}$ used, since the range where the logarithms of $r$ are summed is different. This error also measures the fact that the data points have some error. 
For the data set I, we find the largest difference between fits with different $r_{ref}$ to be of order 8 MeV. For the other data sets the spread is slightly smaller, except for the data set IV, which is slightly larger ($\sim 9$ MeV) after considering the most extreme difference. This one is obtained with the largest $r_{ref}$ we have in our data set, which, on the other hand, produces a rather large $\chi^2_{red}$: $\chi^2_{red}\simeq 4.8$.

{\bf Dependence on $Z_{3}^F$}\\
We have also studied the dependence of our central value on $Z_{3}^F$. We find it to be very small compared with other uncertainties, since the contribution associated to $\Omega_3^F$ is small for our central value determinations. The variation does not change the last digit. Therefore, we will omit it for the final error budget. It is worth mentioning though that for other values of $\nu_s$ and $\nu_{us}$ the terminant is important and, when so, it is a crucial element to get agreement with our central value. 

{\bf Estimate of  higher order contributions}\\
For the error analysis, we need to determine the error associated to our lack of knowledge of the complete perturbative series. We have several ways to estimate this error. We have studied the error produced by the variation of $\nu_s$ and find it to be very small, of the order of 2 MeV. We have also studied the error produced by the variation of $\nu_{us}$ and find it to be of around 9 MeV for the data set I. As an alternative way to estimate the error, we considered the difference between the NNNLL and NNNLL$_{hyp}$, i.e. adding or subtracting the terminant. This produces a very small shift. Alternatively, we have also performed fits changing the order at which we start including the terminant in the force from three to two. We indeed find the variation to be small: $\sim 6$ MeV. The fits have been performed using the running of $\alpha$ with 4 loop accuracy \cite{vanRitbergen:1997va}, as it is the analogous accuracy to the perturbative expansion of the static energy. We have also made the fit including the running of $\alpha$ with 5 loop accuracy \cite{Baikov:2016tgj}, and find a 3 MeV difference with our central value. To consider more conservative estimates of the error, we have also looked at the difference between NNLL and NNNLL fits. For the data set I, we obtain similar numbers, marginally larger, than from the variation in $\nu_{us}$: $\sim 10$ MeV. We take the largest of all these possibilities. We believe this yields a conservative error estimate for the higher order contributions. 

{\bf Final numbers}\\
 Out of this analysis, we proceed to give our prediction, for which we use the data set I. It reads
\be
\label{eq:LambdacentralI}
\Lambda_{\MS}^{(n_f=3)}=338(2)_{\rm stat}(10)_{\rm h.o.}(8)_{\rm r_{ref}}\; {\rm MeV}
\,.
\ee
The central value is taken from the fit of the NNNLL$_{hyp}$ theory expression with $(\nu_s,\nu_{us})=(1/r,C_A\al(\nu_s)/(2r))$ to the data Set I. The first error is the statistical error of the fit. The second one is the one associated to higher order corrections. We estimate it by taking the biggest number among the different estimates for higher order corrections we have discussed above, which corresponds to the difference between the NNNLL and NNLL number. We finally consider doing the fit with different $r_{ref}$. We take the largest difference. This error is a mixture of two sources: on the one hand it is partially related to our lack of knowledge of higher order logarithms, and, on the other, on the error of the lattice data point. Still, we will treat it as an additional source of error. 
We then combine all errors in quadrature and obtain
\be
\label{eq:LambdaFinal}
\Lambda_{\MS}^{(n_f=3)}=338(12)\; {\rm MeV} \,.
\ee
Note that the determinations of $\Lambda_{\MS}^{(n_f=3)}$ obtained at LL, NLL, NNLL and NNNLL are all perfectly inside the one sigma error bar quoted in \eq{eq:LambdaFinal}, as you can see in Fig. \ref{fig:FitUS}. We also give the strong  coupling constant at the scale of $M_{\tau}$ it corresponds to. We obtain
\be
\label{eq:alphaFinalnf3}
\alpha^{(n_f=3)}(M_{\tau})=0.3151(65)\,.
\ee
This number can be compared with other determinations of the strong coupling at around these low energies. One can, for instance, compare with determinations using the heavy quarkonium spectrum \cite{Mateu:2017hlz,Peset:2018ria}. Those also have as a fundamental input the static potential but, at present, they suffer from larger errors than those presented here. In this respect, applying hyperasymptotic expansions to these analyses may improve the accuracy of such determinations. 

Out of these numbers we can also determine $\alpha^{(n_f=5)}(M_z)$. We follow the preferred method advocated in \cite{Herren:2017osy}, which has built in the error from decoupling and truncation when going from the scales we have made the fit up to the $M_z$ mass. We obtain
\be
\label{eq:alphaFinal}
\alpha^{(n_f=5)}(M_z)=0.1181(8)_{\Lambda_{\MS}}(4)_{M_{\tau} \rightarrow M_z}=0.1181(9)\,.
\ee
The first error is the error associated to the error of our determination of $\Lambda_{\MS}^{(n_f=3)}$, and the second to the transformation of this number to $\alpha^{(n_f=5)}(M_z)$ as described in \cite{Herren:2017osy}. In the last equality we have combined the errors in quadrature. Our number is perfectly consistent with the world average number \cite{Tanabashi:2018oca}, or with the lattice final FLAG average value \cite{Aoki:2019cca}, and with a very competitive error. 

As we have mentioned above, our prediction has been obtained using the data set I, which is the one less sensitive to long distances. Nevertheless, we have performed similar error analyses for the other data sets. We find 
\bea
\label{eq:LambdacentralII}
{\rm Set\; II}& \quad& \Lambda_{\MS}^{(n_f=3)}=341(1)_{\rm stat}(11)_{\rm h.o.}(6)_{\rm r_{ref}}\; {\rm MeV}=341(14)\; {\rm MeV}
\,,
\\
{\rm Set\; III}& \quad& \Lambda_{\MS}^{(n_f=3)}=343(1)_{\rm stat}(13)_{\rm h.o.}(7)_{\rm r_{ref}}\; {\rm MeV}=343(14)\; {\rm MeV}
\,,
\\
{\rm Set\; IV}& \quad& \Lambda_{\MS}^{(n_f=3)}=343(0)_{\rm stat}(13)_{\rm h.o.}(9)_{\rm r_{ref}}\; {\rm MeV}=343(16)\; {\rm MeV}
\,.
\eea
Notice that all the central values obtained with the different data sets are within one sigma of our preferred value. The data sets II, III, IV have smaller statistical errors, but larger errors associated to higher order in perturbation theory effects, as they suffer from a larger difference between the NNLL and NNNLL result. 

\begin{figure}[htb] 
%\unitlength=1mm
\centering\includegraphics[width=140mm]{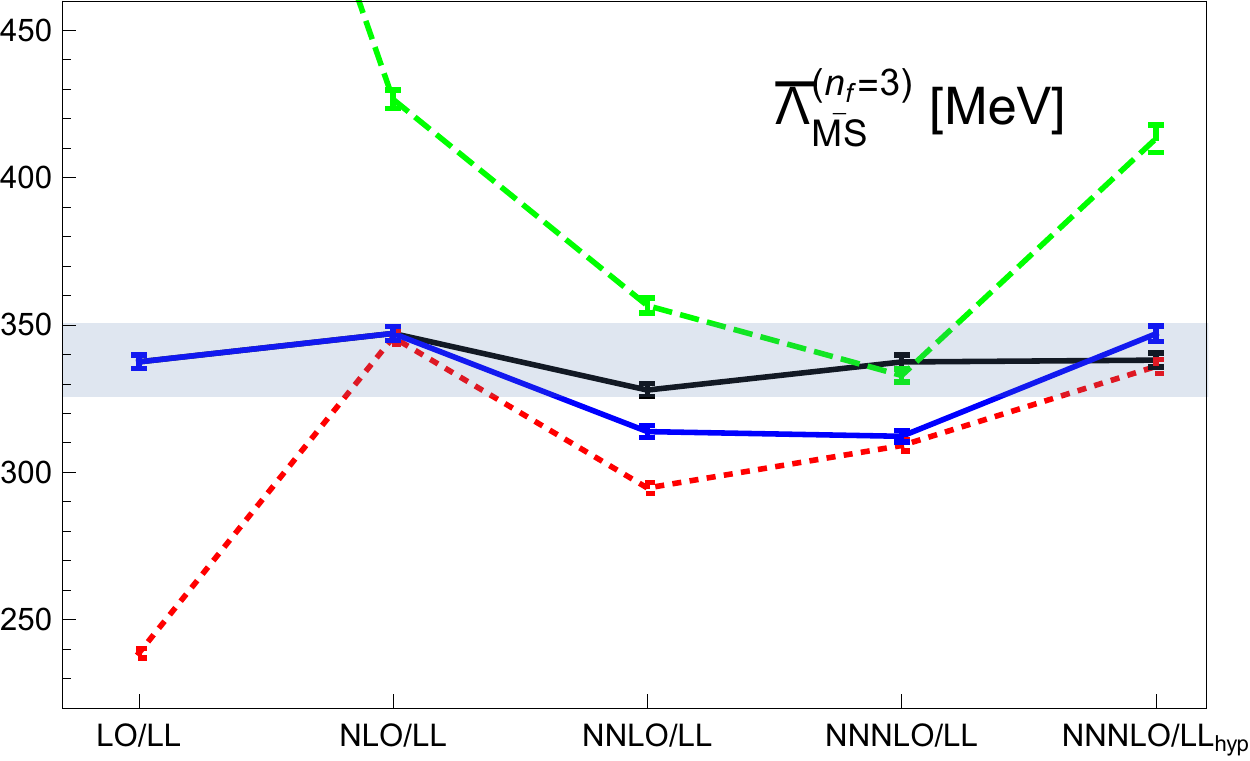}
  % \centering\includegraphics[width=140mm]{FigFO.pdf}
%\vspace{-0.2cm}
\caption{Determination of $\Lambda_{\MS}^{(n_f=3)}$ at LL, NLL, NNLL, NNNLL and NNNLL$_{hyp}$ using the data set I with $(\nu_s,\nu_{us})=(1/r,C_A\al(\nu_s)/(2r))$ (continuous black line). 
We also give the determination of $\Lambda_{\MS}^{(n_f=3)}$ at LO(LL), NLO(NLL), NNLO, NNNLO and NNNLO$_{hyp}$ using the data set I with $\nu_s=\nu_{us}=2/r$ (dashed green line), $\nu_s=\nu_{us}=1/r$ (continuous blue line) and $\nu_s=\nu_{us}=1/(\sqrt{2}r)$ (dotted red line). The error displayed is only the statistical error of the fits. We also show the error band generated by our prediction \eq{eq:LambdaFinal}.} 
\label{fig:FitFO}
\end{figure}

{\bf Comparison with fixed order computations} \\
Fixed order computations can be obtained from the RG improved ones by setting $\nu_s=\nu_{us}$. Therefore, this approximation does not incorporate the resummation of large ultrasoft logarithms. This effect can be important. We show the results of the fixed order computation and the comparison with 
the RG improved result in Fig. \ref{fig:FitFO}. Let us first remind that the first two orders are equal, i.e.: LL=LO and NLO=NLL, as there are no ultrasoft logarithms to resum. The difference shows up at higher orders. For the same value of $\nu_s$, and for order NNLO and NNNLO versus NNLL and NNNLL, the fits at fixed order give significantly lower values than those that perform the resummation of logarithms. On top of that, the incorporation of the $u=3/2$ terminants does not improve the convergence of the determination, unlike when resuming the large ultrasoft logarithms, where we see a very nice convergence pattern. This shows that the resummation of large logarithms appears to be compulsory to find convergence, and to cancel the scale dependence that we have in the terminants. The magnitude of the incorporation of the terminants is larger for larger $\nu_s$. This may say that using $N=0$ for the ultrasoft contribution for a scale as large as $\nu_{us}=2/r$ could be a bad approximation. In this respect notice that as we lower $\nu_{us}$, (see the fits with $\nu_{us}=1/r$, and, particularly, with $\nu_{us}=1/(\sqrt{2}r)$ in Fig. \ref{fig:FitFO}), the convergence of the fixed order computation significantly improves. 

{\bf What about if $\nu_s=$ constant?}\\
In principle, the optimal way to resum the large logarithms is to scale $\nu_s$ with $1/r$ and $\nu_{us}$ with $\al(\nu_s)/r$. In practice, the range of scales we have is not that large. We then consider fits with fixed $\nu_s$ and $\nu_{us}$.  
We choose $(\nu_s,\nu_{us})=(1/r_{ref}, 1$ GeV). For the data sets I, II, III and IV, the NNNLL$_{hyp}$ fits yield 
$\Lambda_{\MS}^{(n_f=3)}=(339,342,344,346)$ MeV respectively. Note that, for the data sets I and II, the result is identical (difference is indeed below 1 MeV and only gets to 1 MeV after rounding) to the central values obtained before and displayed in \eq{eq:LambdacentralI} and \eq{eq:LambdacentralII} respectively. For the data set III, the difference is 1 MeV, and for the data set IV, the difference is slightly more significant:  3 MeV. This agreement is very rewarding, since fits at low orders in the hyperasymptotic approximation show large differences with the analogous fits using the default scales: $(\nu_s,\nu_{us})=(1/r,C_A\al(\nu_s)/(2r))$. For the data set I, we show the values of $\Lambda_{\MS}^{(n_f=3)}$ obtained with $(\nu_s,\nu_{us})=(1/r_{ref}, 1$ GeV) (i.e. the RG improved results) and with $\nu_s=\nu_{us}=1/r_{ref}$ (i.e. the fixed order results) in Fig. \ref{fig:Figpotential}. For the RG improved results we observe how the LL, NLL are outside the error band (actually the LL fit have a large $\chi^2_{\rm red} \simeq 3.9$, which then goes below 1 as we increase the accuracy) but then steadily converge to the central value, such that, as we said, the difference for the NNNLL$_{hyp}$ prediction is below 1 MeV. The fixed order fits, which are also displayed in Fig. \ref{fig:Figpotential}, show the same kind of behavior to the one discussed in the previous item.

{\bf Nonperturbative corrections}\\
In all the determinations above, we have assumed that the ultrasoft scale is in the perturbative regime. In this situation, nonperturbative effects are parametrically suppressed compared with the precision obtained with our hyperasymptotic approximation. This assumption is safer if we take the points with higher energy. For the points of the data Set I, $\nu_{us}$ moves in the range $\nu_{us}=\frac{C_A\al(\nu_s)}{2r} \in (1.06,0.86)$ GeV, for which we consider safe to use perturbation theory at the ultrasoft scale. 

If the ultrasoft scale is in the nonperturbative regime, we can say little from first principles about $\frac{d}{dr}\delta E_{us}$. To make an estimate, we consider the data set IV after subtracting the points of the data set I (those at smallest distances that we used in the previous section for the determination of $\Lambda_{\MS}$ in the purely perturbative regime). As a test, we assume that for this set of data the ultrasoft scale is in the nonperturbative regime. To simplify the parameterization of these nonperturbative effects, we assume that we are in the regime where $1/r \gg \lQ \gg \al(1/r)/r$. In this situation, $\delta E_{us}^{\rm PV}=k_{\rm PV} \Lambda_{\MS}^3 r^2$ (where $k_{\rm PV}$ is a nonperturbative dimensionless constant), instead of being equal to \eq{eq:deltaEUS} plus the terminant (i.e., the $-\frac{1}{r^2}\Omega_3^{F}(\nu_{us})$ is not included in the fit, unlike in the pure perturbative case, as such contribution is inside the nonperturbative term). We first want to see how sensitive the determination of $\Lambda_{\MS}$ would be to considering the ultrasoft scale to be in the nonperturbative regime, which implies that we also have to fit $\delta E_{us}^{\rm PV}$. For such fit, we obtain $\Lambda_{\MS}=356(3)$ 
%0.00865327
MeV (the error is only the statistical error of the fit) with a $\chi^2_{\rm red}=0.55$ (in the fit we have fixed $\nu_{us}=1$ GeV in $V_{\rm PV}^{\rm RG}$). Notice that this number for $\Lambda_{\MS}$ is consistent with the value obtained from the pure perturbative fit (only a little bit more than one sigma away of \eq{eq:LambdaFinal}). For the value of $k_{\rm PV}$ we obtain 
\be
\label{kPVNP}
k_{\rm PV}=-0.82(7)
\,.
\ee
In principle, we did not care much about $k_{\rm PV}$. Nevertheless, this value of $k_{\rm PV}$ rings a bell. In the perturbative regime we have that 
\be
\delta E_{us}^{\rm PV}=\sum_{n=0}^N c_n \alpha^{n+1}(\nu_{us})-\frac{1}{r}\Omega^V_3(\nu_{us})
\,,
\ee
where we set $N=0$. At low scales this expression is dominated by the terminant, which, we remind, has the following form
\be
-\frac{1}{r}\Omega^V_3(\nu_{us}) = k\sqrt{\alpha(\nu_{us})}\Lambda_{\MS}^3 r^2(1+k'\alpha(\nu_{us})+{\cal O}(\alpha^2(\nu_{us}))
\,.
\ee
The dependence on $\nu_{us}$ is mild and, effectively, the terminant scales as 
\be
-\frac{1}{r}\Omega^V_3(\nu_{us}) 
\sim k^{\rm terminant}_{\rm PV}\Lambda_{\MS}^3 r^2
\,,
\ee
and for $\nu_{us}=1$ GeV we get
\be
\label{kPVterminant}
k^{\rm terminant}_{\rm PV}\simeq -1.25(58)\;, \quad k_{\rm PV}^{\rm terminant} \simeq -1.04(48) 
\,,
\ee
where in the first number we have used $\Lambda_{\MS}^{(n_f=3)}=338$ MeV, the outcome of the perturbative fit, and in the second $\Lambda_{\MS}^{(n_f=3)}=356$ MeV, the outcome of the nonperturbative fit. 
In these numbers we have put the central value and the error of the normalization $Z_3^V$ obtained in \eq{ZF3nf3}. Therefore, what the nonperturbative fit does is to effectively fit the terminant assuming that the ${\cal O}(\al(\nu_{us}))$ term of $\delta E_{us}^{\rm PV}$ is subdominant. Notice that \eq{kPVNP} is, within one statistical standard deviation, the value predicted by perturbation theory, \eq{kPVterminant}. We take this as a very strong confirmation that our weak coupling analysis is safe, and that, indeed, one can apply perturbation theory to scales as small as $1/r \sim 1$ GeV. Finally, to confirm this picture, we do the fit over the complete data set IV assuming $\delta E_{us}^{\rm PV}=k_{\rm PV} \Lambda_{\MS}^3 r^2$. The results barely change: we obtain $\Lambda_{\MS}=355(3)$ with also $\chi^2_{\rm red}=0.55$ and $k_{\rm PV}=-0.8$. Overall, we can even take this analysis as a strong indication that the data has enough precision to be sensitive (and, to some extent fit, albeit with large errors) the value of $Z_3^V$.
This discussion also explains why the nonperturbative fit also has a small $\chi^2_{\rm red}$, as it loosely corresponds to the perturbative expression but letting the normalization of the terminant to be a free parameter of the fit. 

{\bf Comparison with earlier work}\\
Determinations of $\Lambda_{\MS}^{(n_f=3)}$ using lattice data of the static energy have been obtained in the past. Some recent determinations are those of \cite{Takaura:2018lpw,Takaura:2018vcy}. They compare with a different data set including lattice data at longer distances. They work directly with the potential. The precision of the theoretical expression is NNNLO in our counting. No resummation of ultrasoft logarithms nor the incorporation of the terminants is considered, but they use an alternative method for dealing with the renormalons. In the large $\beta_0$, it is possible to see what is the precision that corresponds to in the hyperasymptotic approximation but not beyond the large $\beta_0$. The ultrasoft scale is assumed to be in the nonperturbative situation. Therefore, the comparison should better be done with the number we have just obtained in the previous item. If we set $\delta E_{us}^{\rm PV}=k_{\rm PV} \Lambda_{\MS}^3 r^2$ and fix $\nu_s=\nu_{us}=1/r$ (i.e. we work with NNNLO$_{hyp}$ precision, we obtain $\Lambda_{\MS}^{(n_f=3)}=305(2)$ MeV where we only put the statistical error. This number is smaller than the number obtained in \cite{Takaura:2018lpw,Takaura:2018vcy}.

Closer to our analysis are \cite{Bazavov:2014soa,Bazavov:2019qoo}. In particular from the last reference we borrow the lattice data. In these references, they use the force as the starting point and later integrate it to recover the potential, as we have done above. Their central values are obtained by fitting to the NNNLO result after adding the NNLL ultrasoft contribution. Therefore, they mix different orders according to our counting and do not include the complete NNNLL result. The number they obtain is smaller than ours. In this respect, note that our numbers with analogous NNNLO precision are also smaller.  

\subsection{Direct fit from the static potential}
\label{Sec:FitV}
We now present an alternative determination of $\Lambda_{\MS}^{(n_f=3)}$ to the one obtained in the previous section.
As in the previous section, we will mainly work with the data set I, but, in this section, we fit the lattice data to \eq{Edifference} (we also consider here energy differences) using \eq{eq:Esapprox}. In this equation we will set $N_P=1$ by default. This also means that $3N_P=N_{max}=3$ and we reach the next renormalon. Nevertheless, we also explore the impact of choosing different values of $N_P$. The counting of the hyperasymptotic expansion will then be the following: The static potential at tree level corresponds to the LO (which is equal to the LL) approximation. In the hyperasymptotic counting $(D,N)$ it corresponds to (0,0) precision. The static potential at one-loop corresponds to the NLO (which is equal to the NLL) approximation. In the hyperasymptotic counting, it corresponds to (0,1) precision. We then add the terminant associated to the $u=1/2$ renormalon. We name such approximation NLO/NLL$_{hyp1}$. In the hyperasymptotic counting, it corresponds to (1,0) precision. We then add the ${\cal O}(\al^3)$ terms in \eq{eq:Esapprox}. We name such approximation  NNLL$_{hyp1}$ or NNLO$_{hyp1}$ if the ${\cal O}(\al^3)$ contributions of $\delta V_{\rm RG}$ are added or not, respectively. In the hyperasymptotic counting, it corresponds to (1,1) precision. We then add the ${\cal O}(\al^4)$ terms in \eq{eq:Esapprox}. We name such approximation  NNNLL$_{hyp1}$ or NNNLO$_{hyp1}$ if the ${\cal O}(\al^4)$ contributions of $\delta V_{\rm RG}$ are added or not, respectively. In the hyperasymptotic counting, it corresponds to (1,2) precision. Finally, we add the terminants associated to the $u=3/2$ renormalon of $V$ and $\delta E_{us}$. Note that each terminant depends on the order we truncate each perturbative series and the scale of $\alpha$ in each respective perturbative expansion. We will name such approximation  NNNLL$_{hyp2}$ or NNNLO$_{hyp2}$ if the ${\cal O}(\al^4)$ contributions of $\delta V_{\rm RG}$ are added or not, respectively. In the hyperasymptotic counting, it corresponds to (3,0) precision.

\begin{figure}[htb] 
%\unitlength=1mm
\centering\includegraphics[width=140mm]{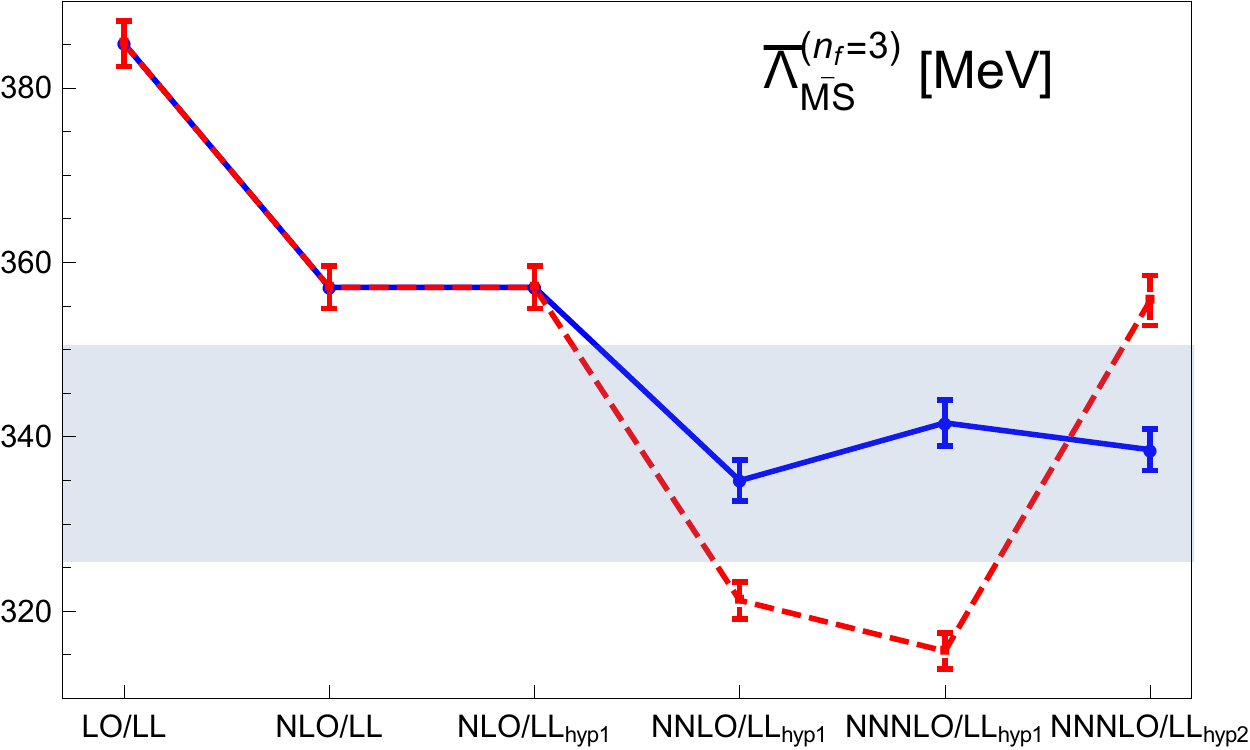}
   %\centering\includegraphics[width=140mm]{Figpotential.pdf}
%\vspace{-0.2cm}
\caption{Determination of $\Lambda_{\MS}^{(n_f=3)}$ at LL, NLL=NLL$_{hyp1}$, NNLL=NNLL$_{hyp1}$, NNNLL=NNNLL$_{hyp1}$, and NNNLL$_{hyp}$=NNNLL$_{hyp2}$ using the data set I with $(\nu_s,\nu_{us})=(1/r_{ref},1$ GeV) (blue continuous line) and with $(\nu_s,\nu_{us})=(1/r_{ref},1/r_{ref}$) (red dashed line)
using \eq{eq:Esapprox} or \eq{Eforceth} (the result is the same). Note that the case with $(\nu_s,\nu_{us})=(1/r_{ref},1/r_{ref}$) is equivalent to 
LO, NLO=NLO$_{hyp1}$, NNLO=NNLO$_{hyp1}$, NNNLO=NNNLO$_{hyp1}$,  NNNLO$_{hyp}$=NNNLO$_{hyp2}$. The error displayed is only the statistical error of the fits. 
We also show the error band generated by our prediction in \eq{eq:LambdaFinal}.} 
\label{fig:Figpotential}
\end{figure}

We will use the fits performed in this section to reassure the results obtained in the previous section. A priori one would expect the error of this determination to be larger due to the error associated to the first renormalon. We can avoid this error completely if we set $\nu_s=1/r_{ref}$. We then first perform a fit setting $\nu_s=\nu_{us}=1/r_{ref}$, so we avoid any $r$ dependence in the renormalization scale. Since 
$\nu_s=1/r_{ref}$, the renormalon associated to $u=1/2$ exactly cancels, since the scale is the same and the perturbative series is truncated at the same order. Therefore, the associated terminants are set to zero. Thus, the computation is equivalent to standard perturbation theory, and, in the counting above, we have NLO=NLO$_{hyp1}$, NNLO=NNLO$_{hyp1}$, NNNLO=NNNLO$_{hyp1}$. As we said, we also put $\nu_s=\nu_{us}$. Nevertheless, in this case, the renormalon associated to $u=3/2$ of the static potential and of $\delta E_{us}$ do not cancel each other. The reason is that the perturbative expansion of the static potential is truncated at $N=3$, whereas the perturbative expansion of $\delta E_{us}$ is truncated at $N=0$. Therefore, the associated terminants do not cancel each other. The results are identical to those found in the previous section when setting $\nu_s$ and $\nu_{us}$ constant. 
We show the results in Fig. \ref{fig:Figpotential}. We then take $\nu_{us}=1$ GeV. This still avoids any $r$ dependence in the renormalization scale and still NLL=NLL$_{hyp1}$, but the following orders are different, but only by a constant that cancels in the energy difference, except for the NNNLL$_{hyp2}$. Again, the results are identical to those found in the previous section using ${\cal F}$ when setting $\nu_s$ and $\nu_{us}$ constant. Therefore, we reach to the same conclusions: We saw in the previous section using ${\cal F}$ that working with $\nu_s=1/r_{ref}$ and $\nu_{us}=1$ GeV produced very similar fits to those with $\nu_s=1/r$ and $\nu_{us}=\frac{C_A\al(\nu_s)}{2r}$ GeV at high orders in the hyperasymptotic expansion, with a difference less than 1 MeV. As the results are identical, we observe the same behavior here: using directly \eq{eq:Esapprox} at NNNLL$_{hyp2}$ with $(\nu_s,\nu_{us})=(1/r_{ref},1$ GeV) yields the same result (with a difference smaller than 1 MeV) than a fit using \eq{Eforceth} at NNNLL$_{hyp}$ with 
$(\nu_s,\nu_{us})=(1/r,\frac{C_A\al(\nu_s)}{2r})$. We show the results in Fig. \ref{fig:Figpotential}. It is also very rewarding to see the stability of this result to changing $N_P$. This far we have set $N_P=1$, but even if we set $N_P=3$ (such that we do not include the subleading renormalon but only the leading one), the fit yields a very similar number for $\Lambda_{\MS}^{(n_f=3)}$: $\Lambda_{\MS}^{(n_f=3)}=342$ MeV. Only four MeV away from our central value. 

\begin{figure}[htb] 
%\unitlength=1mm
%\centering\includegraphics[width=120mm]{Fig3.pdf}
   \centering\includegraphics[width=140mm]{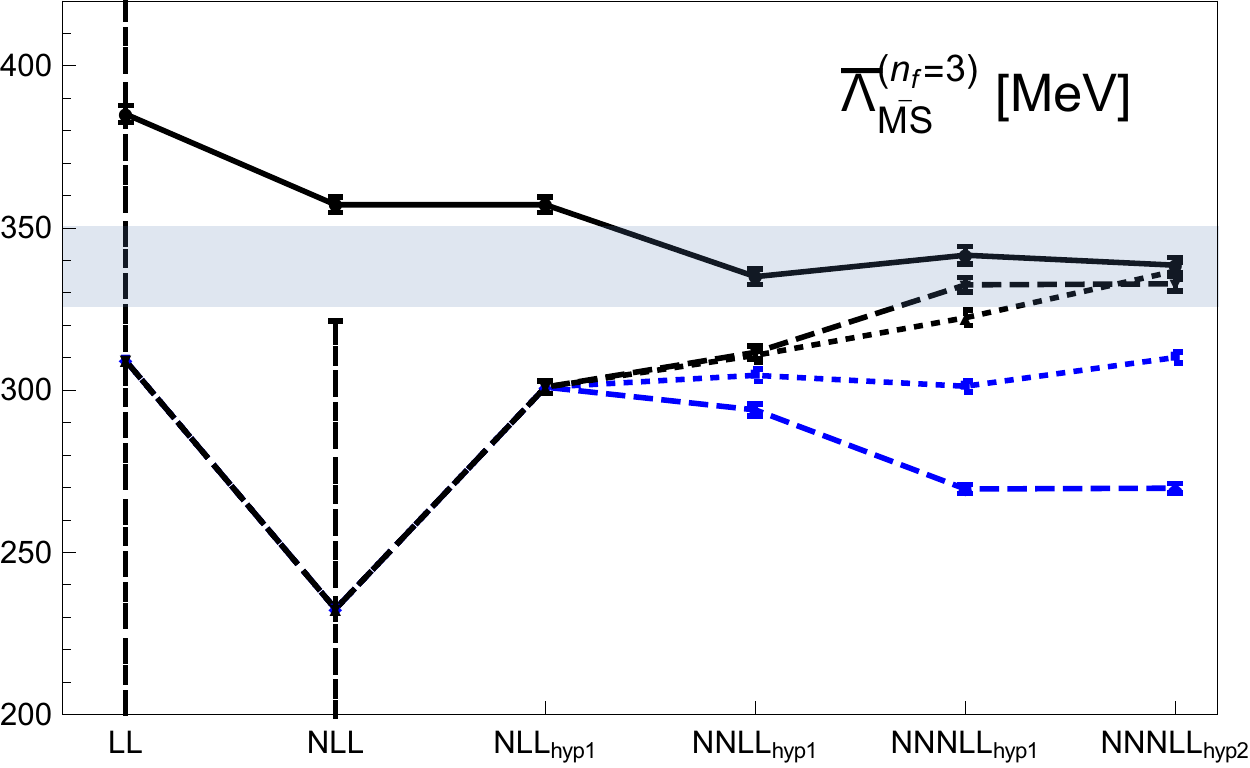}
%\vspace{-0.2cm}
\caption{ Determination of $\Lambda_{\MS}^{(n_f=3)}$ at LL, NLL, NLL$_{hyp1}$, NNLL$_{hyp1}$, NNNLL$_{hyp1}$, and NNNLL$_{hyp2}$ using the data set I with $(\nu_s,\nu_{us})=(1/r_{ref},1$ GeV) (continuous black line), with $(\nu_s,\nu_{us})=(1/r,1$ GeV) (dotted blue line), and with $(\nu_s,\nu_{us})=(1/r,\frac{C_A \al(\nu_s)}{2r}$) (dashed blue line) using \eq{eq:Esapprox}. We also show the determinations of $\Lambda_{\MS}^{(n_f=3)}$ at LL, NLL, NLL$_{hyp1}$, NNLL$_{hyp1}$, NNNLL$_{hyp1}$, and NNNLL$_{hyp2}$ using the data set I with $(\nu_s,\nu_{us})=(1/r,1$ GeV) (dotted black line), and with $(\nu_s,\nu_{us})=(1/r,\frac{C_A \al(\nu_s)}{2r}$) (dashed black line) using \eq{eq:Esapprox} except for $\delta V_{RG}$, for which we use \eq{Eforceth}. 
The error displayed for the fits is only the statistical error.
We also show the complete error band generated by our prediction \eq{eq:LambdaFinal}.} 
\label{fig:FigpotentialdeltaVRG}
\end{figure}

We now introduce the $r$ dependence. Surprising things happen. Working with $\nu_s=1/r$ or with $\nu_s=$constant show a qualitative different behavior. The reason was first explained in \cite{Pineda:2002se}, and it is due to the different ways the behavior of the leading renormalon appears at finite orders in perturbation theory. Working with $\nu_s=$constant, the renormalon-associated contribution to the fixed order term of the perturbative expansion is $r$ independent, but this is not so for $\nu_s=1/r$. In this last case, one has to be much more careful in dealing with the renormalon and to enforce its cancellation. What we see is that even at ${\cal O}(\alpha)$ the leading renormalon plays a very important role. We get a very high $\chi^2_{red}= 3650$. At ${\cal O}(\al^2)$ the fit is also bad with a $\chi^2_{red}= 2856$. This should be compared with the $\chi^2_{red}$ obtained when $\nu_s=1/r_{ref}$ ($\chi^2_{red}= 3.9$ at ${\cal O}(\al)$ and $\chi^2_{red}=0.9$ at ${\cal O}(\al^2)$). It is only when the terminant associated to the leading renormalon is included at NLO$_{hyp1}$ that the fit is reasonable and yields a $\chi_{red}=0.4$, below 1. This means that we have already reached the asymptotic behavior with $N_P=1$ or rather with $N_P=0$. We show the results in Fig. \ref{fig:FigpotentialdeltaVRG}. We next consider what happens if we also introduce $r$ dependence in $\nu_{us}$ by setting $\nu_{us}=C_A\al(\nu_s)/(2r)$. The situation, in this case, is even worse. We also show the results in Fig. \ref{fig:FigpotentialdeltaVRG}. 

We have investigated the origin of the problem. The fact that it only shows up when we introduce $r$ dependence in $\nu_s$ and $\nu_{us}$ induces to think that it has to do with renormalons, similarly to the discussion one can find in \cite{Pineda:2002se} for the leading renormalon. The leading renormalon is under control. Therefore, the issues we face should have to do with subleading renormalons, maybe also with the renormalons encoded in $\Delta V$. Here we do not have a clear explanation. We postpone a detailed analysis to future work. What we have been able to do is to identify where the effect seems to be hidden, and it is in $\delta V_{\rm RG}$. If instead of using \eq{eq:deltaVRG}, we obtain $\delta V_{\rm RG}$ from the perturbative expansion of its derivative, \eq{eq:derivativedeltaVRG}, which we then integrate, we find that most, if not all, of the difference cancels. We show this effect in Fig. \ref{fig:FigpotentialdeltaVRG}. We will not perform a full-fledged error analysis in this case, though, as we do not have a clear understanding of how the subleading renormalons are showing up when $\nu_s=1/r$ and $\nu_{us}=C_A\al(\nu_s)/(2r)$. Finally, even though it is not discussed in this paper, notice that, when working with $\nu_s=1/r$, the error associated to $Z_1^V$ shows up, and it can be potentially large. 

\section{Conclusion}

In this paper, we have obtained hyperasymptotic approximations for the static energy and for the force with a precision of $(D,N)=(3,0)$. Our expressions also implement the resummation of large logarithms to NNNLL precision. We have used these expressions to obtain very precise determinations of $\Lambda_{\MS}$ and $\al^{(n_f=5)}(M_z)$. We have mainly used the force as the starting point of our theoretical analyses.  Our final result reads
\be
\label{eq:LambdaFinalCon}
\Lambda_{\MS}^{(n_f=3)}=338(12)\; {\rm MeV} \,, \quad \alpha(M_{\tau})=0.3151(65) \,,
\quad \alpha(M_z)=0.1181(9)\,.
\ee
The resummation of logarithms and the introduction of the terminants associated to the $u=3/2$ renormalon are essential to get a very well convergent series. This, together with precise data at short distances, allows us to get accurate values for $\Lambda_{\MS}$. The lack of any of these novel elements significantly deteriorates the convergence, and consequently, the accuracy of the prediction. 

Our fits are based on the shortest available data from \cite{Bazavov:2017dsy,Bazavov:2019qoo} that do not suffer from lattice artifacts. This means that we have restricted the fit to $r$ smaller than 0.353 GeV$^{-1}$. To make the weak coupling analysis as reliable as possible, the smallest soft scale we take is 2 GeV. This corresponds to an ultrasoft scale of 0.86 GeV. The fit shows no signal of needing extra nonperturbative correction. This is so even if we relax the infrared cutoff of the soft scale down to 1 GeV. The fits are still perfectly ok. In this respect, we do not have any indication from the fit of the need of nonperturbative corrections. One strong, very nontrivial, validation that we can use perturbation theory in our fit actually comes from fits assuming that the ultrasoft scale is in the nonperturbative regime. We do fits assuming that we are in the situation where $\Lambda_{QCD} \gg \al(\nu_s)/r$. In this energy regime $\delta E_{us}^{\rm PV}=k_{\rm PV} \Lambda_{\MS}^3 r^2$ and, besides, $\Lambda_{\MS}$, we also have to fit $k_{\rm PV}$. Within one sigma of the statistical error of the fit, the number obtained for $k_{\rm PV}$ is the same as the one predicted by the terminant of the weak coupling expression of $\delta E_{us}^{\rm PV}$. 

The largest source of error comes from unknown higher order corrections in perturbation theory. The statistical errors of the fit are small, though the dependence on $r_{ref}$, which is a mixture of lattice and theory error is large. Increasing the number of points of the data set gives a very mild tendency to increase the value till stabilizing at 343 MeV, very well inside the error we give.  

Whereas the cancellation of the leading renormalons is under control, the situation is not that clear for subleading renormalons. $\delta E_{us}$ depends on $\Delta V$, which has a renormalon located at $u=1/2$. The specific way this renormalon cancels is something that should be investigated. This is what has stopped us from using the perturbative NLO expression for $\delta E_{us}$. In this respect, the analysis of \cite{Pineda:2010mb} could be of help. In this reference, linear power-like divergences that appear in the coefficients of the perturbative expansion due to renormalons located at $u=1/2$ become logarithmic divergences that are identified in dimensional regularization. In that specific example, one could see the cancellation between different terms. Likely related with this discussion, there is another issue that has to be investigated: the existence of renormalons in $\delta V_{\rm RG}$. We observe a rather different behavior if we first derive it and afterwards integrate it or if we directly 
use \eq{eq:deltaVRG}. This difference only appears if $\nu_s$ and $\nu_{us}$ are made to be explicitly $r$ dependent. This is consistent with the existence of a renormalon in this object. Indeed, it resembles the situation faced by early studies of the leading renormalon \cite{Pineda:2002se}. In that case, depending on how one does the expansion, in powers of $\al(1/r)$ or in powers of $\al(\nu_s={\rm constant})$, the perturbative series was also convergent or not. Working with the force, we observe that the cancellation of renormalons is incorporated from the start, irrespectively of working with $\nu_s=$constant or not. In the case of working with the potential, we are, at present, only safe if working with $\nu_s$ and $\nu_{us}$ being $r$-independent. These issues will be investigated in future work. 

\medskip
 
{\bf Acknowledgments}\\
\noindent
We thank J. H. Weber for sharing with us the lattice data of \cite{Bazavov:2019qoo}, for several clarifying remarks, and for comments on the manuscript.
We thank P. Petreczky for sharing with us the lattice data of \cite{Bazavov:2017dsy}, for clarifying remarks, and for comments on the manuscript. We also thank O. Kaczmarek for the lattice data of \cite{Cheng:2007jq}. We thank X. Garcia i Tormo for clarifying remarks with respect \cite{Bazavov:2014soa}. We thank N. Brambilla and A. Vairo for discussions. We thank A. Kataev for references \cite{Gorishnii:1991hw,Lee:2016cgz}.  C.A. thanks the IFAE group at Universitat Aut\`onoma de
Barcelona for warm hospitality during part of this work. 
This work was supported in part by the Spanish grants FPA2017-86989-P and SEV-2016-0588 from the ministerio de Ciencia, Innovaci\'on y Universidades, and the grant 2017SGR1069 from the Generalitat de Catalunya; and 
by FONDECYT (Chile) under grant No. 1200189. This project has received funding from the European Union's Horizon 2020 research and innovation programme under grant agreement No 824093.

\appendix
\section{Constants}
\label{constants}

The coefficients $a_n$ we define in \eq{eq:Vn} read ($a_0(\nu_s r;\frac{\nu_{us}}{\nu_s})=1$) 
\begin{eqnarray}
a_1(\nu_s r)
&=&
a_1+2\beta_0\,\ln\left(\nu_s e^{\gamma_E} r\right)
\,,
\nonumber\\
a_2(\nu_s r)
&=&
a_2 + \frac{\pi^2}{3}\beta_0^{\,2}
+\left(\,4a_1\beta_0+2\beta_1 \right)\,\ln\left(\nu_s e^{\gamma_E} r\right)\,
+4\beta_0^{\,2}\,\ln^2\left(\nu_s e^{\gamma_E} r\right)\,
\,,
\nonumber \\
a_3(\nu_s r;\frac{\nu_{us}}{\nu_s})
&=&
a_3+ a_1\beta_0^{\,2} \pi^2+\frac{5\pi^2}{6}\beta_0\beta_1 +16\zeta_3\beta_0^{\,3}
\nonumber \\
&+&\bigg(2\pi^2\beta_0^{\,3} + 6a_2\beta_0+4a_1\beta_1+2\beta_2\bigg)\,
  \ln\left(\nu_s e^{\gamma_E} r\right)+\frac{16}{3}C_A^{\,3}\pi^2\,
  \ln\left(\nu_{us} e^{\gamma_E} r\right)\,
\nonumber \\
&+&\bigg(12a_1\beta_0^{\,2}+10\beta_0\beta_1\bigg)\,
  \ln^2\left(\nu_s e^{\gamma_E} r\right)\,
+8\beta_0^{\,3}  \ln^3\left(\nu_s e^{\gamma_E} r\right)\,
\,.
\label{eq:Vr}
\end{eqnarray}
\be
a_1={31C_A-20T_Fn_f \over 9};
\ee
\bea
\nonumber
&a_2 =& 
{{400\,{{{\it n_f}}^2}\,{{{\it T_F}}^2}}\over {81}} -
     {\it C_F}\,{\it n_f}\,{\it T_F}\,
      \left( {{55}\over 3} - 16\,\zeta(3) \right) 
\\
&&
\nn
 +
     {{{\it C_A}}^2}\,\left( {{4343}\over {162}} +
        {{16\,{{\pi }^2} - {{\pi }^4}}\over 4} + {{22\,\zeta(3)}\over 3}
        \right)
 - {\it C_A}\,{\it n_f}\,{\it T_F}\,
      \left( {{1798}\over {81}} + {{56\,\zeta(3)}\over 3} \right)
	  ;
\eea
\begin{eqnarray}
  a_3 = a_3^{(3)} n_f^3 + a_3^{(2)} n_f^2 + a_3^{(1)} n_f + a_3^{(0)}
  \,,
\end{eqnarray}
where
\begin{eqnarray}
  a_3^{(3)} &=& - \left(\frac{20}{9}\right)^3 T_F^3
  \,,\nonumber\\
  a_3^{(2)} &=&
  \left(\frac{12541}{243}
    + \frac{368\zeta(3)}{3}
    + \frac{64\pi^4}{135}
  \right) C_A T_F^2
  +
  \left(\frac{14002}{81}
    - \frac{416\zeta(3)}{3}
  \right) C_F T_F^2
  \,,\nonumber\\
  a_3^{(1)} &=&
  -709.717 C_A^2 T_F
  +
  \left(-\frac{71281}{162}
    + 264 \zeta(3)
    + 80 \zeta(5)
  \right) C_AC_F T_F
  \nonumber\\&&\mbox{}
  +
  \left(\frac{286}{9}
    + \frac{296\zeta(3)}{3}
    - 160\zeta(5)
  \right) C_F^2 T_F
 -56.83(1) \frac{d_F^{abcd}d_F^{abcd}}{N_A} \,,
  \nonumber\\
  a_3^{(0)} &=&
  502.24(1) \,\, C_A^3
  -136.39(12)\,\, \frac{d_F^{abcd}d_A^{abcd}}{N_A}
  \,,
  \label{eq::a3}
\end{eqnarray}
and
\be
\frac{d_F^{abcd}d_F^{abcd}}{N_A}=
\frac{18-6N_c^2+N_c^4}{96N_c^2}\,,
\quad
\frac{d^{abcd}_F d^{abcd}_A}{N_A}=\frac{N_c(N_c^2+6)}{48}
\,.
\ee
Analytic expressions for $a_3^{(1)}$ and $a_3^{(0)}$ can be found in \cite{Lee:2016cgz}.\footnote{We thank Andrei Kataev for pointing this out.}

\section{${\cal F}^{\rm RG}(r)$ with $\nu_s=x_s/r$ and $\nu_{us}=x_{us}\frac{C_A\al(\nu_s)}{2r}$}
\label{app:nu}
We have (we only discuss the pure perturbative terms)
\bea
{\cal F}^{\rm RG}(r)=F_{\rm PV}(\nu_{us}=\nu_s)
+\frac{d}{dr}\delta V_{s,RG}(r;\nu_s,\nu_{us})+\frac{d}{dr}\delta E_{us}^{\rm PV}(r;\nu_{us})
\,,
\eea
where the coefficients of $F_{\rm PV}(\nu_{us}=\nu_s)$ are 
\bea
\label{eq:fnnew}
f_0(x_s)&=&\frac{C_F}{r^2} \;, \qquad f_1(x_s)=\frac{C_F}{4\pi r^2}\left(a_1(x_s)-2\beta_0\right)\,,
\\
\nn
f_2(x_s)&=&\frac{C_F}{(4\pi)^2 r^2}
\left(
a_2(x_s)-4a_1(x_s)\beta_0-2\beta_1
\right)
\,,
\\
\nn
f_3(x_s;1)&=&\frac{C_F}{(4\pi)^3 r^2}
\left(
a_3(x_s;1)-6a_2(x_s)\beta_0-4a_1(x_s)\beta_1-2\beta_2
\right)
\,.
\eea
Note that these coefficients are equal to those in \eq{eq:fn} with $\nu_{us}=\nu_s$ except for $f_3$. For the other terms we have 
\be
\frac{d}{dr}\delta E_{us}^{\rm PV}(r;\nu_{us})
=
C_F r(\Delta V)^3 
\frac{\al(\nu_{us}) }{9 \pi} \left(6\ln\frac{\Delta V}{\nu_{us}}+6\ln 2-5\right)
\,,
\ee
\bea
\nn
&&
\frac{d}{dr}\delta V_{s,RG}(r;\nu_s,\nu_{us})
=-r(\Delta V)^3
G(\nu_s;\nu_{us})+C_FV_A^2 r(\Delta V)^3 \frac{2\al(\nu_s)}{\pi}\ln \frac{\al(\nu_{us})}{\al(\nu_s)}
\\
&&
+r(\Delta V)^3C_F\frac{2}{3\pi}\left(\al(\nu_{us})-\al(\nu_s)\right)+{\cal O}(\al^5)
\,.
\eea
It is remarkable that the differences among the different terms cancel each other such that the expression for ${\cal F}^{\rm RG}(r)$ we obtain here is equal to the one used in Sec. \ref{Sec:Force}.

\end{document}